\def\nii{[N~{\sc ii}]}
\def\teoiii{T$_{\rm e}$[O III]}
\def\tenii{T$_{\rm e}$[N II]}
\def\teoii{T$_{\rm e}$[O II]}
\def\tesiii{T$_{\rm e}$[S III]}
\def\oiii{[O~{\sc iii}]}
\def\oi{[O~{\sc i}]}
\def\oii{[O~{\sc ii}]}
\def\ha{H$\alpha$}
\def\hb{H$\beta$}

\def\hii{H~{\sc ii}}
\def\heii{He~{\sc ii}}

\def\hei{He~{\sc i}}
\def\sii{[S~{\sc ii}]}
\def\siii{[S~{\sc iii}]}
\def\cbeta{c({H$\beta$})}

\def\nii{[N~{\sc ii}]}

\def\oiii{[O~{\sc iii}]}
\def\oi{[O~{\sc i}]}
\def\oii{[O~{\sc ii}]}
\def\ha{H$\alpha$}
\def\hb{H$\beta$}

\def\hii{H~{\sc ii}}
\def\hi{H~{\sc i}} \def\heii{He~{\sc ii}}

\def\hei{He~{\sc i}}
\def\sii{[S~{\sc ii}]}
\def\siii{[S~{\sc iii}]}
\def\te{$T_{e}$}

\documentclass[structabstract]{aa}  
\usepackage{graphicx}
\usepackage{rotating}
\usepackage{psfig}
\usepackage{txfonts}
\usepackage{longtable,lscape}
\begin{document}

\title{Metal production in M33: space and time variations}
\author{Laura Magrini\inst{1}, 
Letizia Stanghellini\inst{2}, 
Edvige Corbelli\inst{1},  
Daniele Galli\inst{1}, 
Eva Villaver\inst{3}}

\institute{
INAF--Osservatorio Astrofisico di Arcetri, Largo E. Fermi 5, I-50125 Firenze, Italy\\
\email{laura--edvige--galli@arcetri.astro.it}
\and
National Optical Astronomy Observatories, Tucson, AZ 85719, USA\\
\email{lstanghellini@noao.edu}
\and
Universidad Aut\'onoma de Madrid, Departamento de F\'{\i}sica
Te\'orica C-XI, 28049 Madrid, Spain\\
\email{eva.villaver@uam.es}\\
}

\date{}
\abstract
{Nearby galaxies are ideal places to study metallicity gradients in detail and their 
time evolution. }
{We analyse the spatial distribution of metals in M33 using a new sample and the literature  data on \hii\ regions, 
and constrain a model of galactic chemical evolution  with   \hii\ region and planetary nebula (PN) abundances.}
{We consider chemical abundances of  a new sample of \hii\ 
regions complemented  with previous  data sets.
We compared \hii\ region and PN abundances obtained 
with a common set of observations taken at MMT. 
With an updated  theoretical model, we followed
the time evolution of the baryonic components 
and chemical abundances in the disk of M33, assuming that the galaxy is
accreting gas from an external reservoir.} 
{From the sample of \hii\ regions, we find that  {\em i}) the 2D metallicity distribution has an off-centre peak located in 
the southern arm; {\em ii}) the oxygen abundance gradients in the northern and southern sectors, as 
well as in the nearest and farthest sides, are identical within the uncertainties, with slopes around -0.03-4 dex kpc$^{-1}$; 
{\em iii}) bright giant \hii\ regions have a steeper abundance gradient than the other \hii\ regions;
{\em iv}) \hii\ regions and PNe have O/H 
gradients very close  within the errors; 
{\em v}) our updated evolutionary model is able to reproduce the new observational 
constraints, as well as the metallicity gradient and its evolution. }
{Supported by a uniform sample of nebular spectroscopic observations, 
we conclude that {\em i}) the metallicity distribution in M33 is very complex, showing 
a central depression in metallicity probably  due to observational bias;
{\em ii})  the metallicity gradient  in the disk of M33 has a slope of  -0.037$\pm$ 0.009~dex~kpc$^{-1}$ 
in the whole 
radial range up to $\sim$8 kpc, and -0.044$\pm$ 0.009~dex~kpc$^{-1}$ excluding the central kpc;
{\em iii}) there is little  evolution in the slope 
with time  from the epoch of PN progenitor formation to the present.}

\authorrunning{Magrini, L. et al.}
\titlerunning{Metal production in M33}
\maketitle

\section{Introduction}

Understanding the chemical evolution of nearby galaxies has become
more and more interesting in the recent past, given the wealth of new,
detailed data available for their stellar and gaseous components.  M33 (NGC 598) 
is an ideal laboratory for testing chemical evolution models. 
As a nearby blue star-forming galaxy,  with a large angular size (optical size 53\arcmin$\times$83\arcmin,
Holmberg \cite{holmberg58}), and an intermediate inclination ($i$=53$^\circ$),  
it is one of the galaxies observed with the greatest resolution and sensitivity.  
Distance estimates range from 730~kpc  (Brunthaler et al.~\cite{brunthaler05}) to 964~kpc (Bonanos et al.~\cite{bonanos06}). In this paper we adopt 
an average value of 840 kpc  (Freedman et al. \cite{freedman91}).
Many spectroscopic
studies are available, aiming at  determining the physical and chemical properties 
of its \hii\ regions (cf., e.g., Smith~\cite{smith75}, Kwitter \&
Aller~\cite{kwitter81}, V\'{\i}lchez et al.~\cite{vilchez88},
Willner \& Nelson-Patel~\cite{willner02}, Crockett et
al.~\cite{crockett06}, Magrini et al. \cite{magrini07a}, Rosolowsky \& Simon \cite{rs08}, Rubin et al. \cite{rubin08}).
These \hii\ regions  trace  the present
interstellar medium (ISM) and their metallicity is a measure of the star
formation rate (SFR) integrated over the whole galaxy lifetime.
As a result,  the metallicity of \hii\ regions gives interesting constraints
to galactic chemical evolution models.  

In the past, studies of the metallicity gradient of \hii\ regions in M33 
have shown a very steep profile.  First quantitative spectroscopic studies
were carried out by Smith~(\cite{smith75}), Kwitter \&
Aller~(\cite{kwitter81}) and V\'{\i}lchez et
al.~(\cite{vilchez88}). Their observations, limited to the
brightest and largest \hii\ regions,  led to a radial oxygen gradient  about -0.1~dex~kpc$^{-1}$.
Considering the observations by the above researchers, Garnett et al.~(\cite{garnett97}) obtained  an overall O/H
gradient of -0.11$\pm$0.02~dex~kpc$^{-1}$.

With increasing sample sizes,  flatter
gradients have been determined, in
particular by {\em i}) Willner \& Nelson-Patel~(\cite{willner02}) who derive neon
abundances of 25 \hii\ regions from infrared lines, obtaining a Ne/H
gradient of -0.034$\pm$0.015~dex~kpc$^{-1}$; {\em ii}) Crockett et
al.~(\cite{crockett06}) who derive even shallower gradients 
(-0.016$\pm$ 0.017~dex~kpc$^{-1}$ for Ne/H and 
-0.012$\pm$0.011~dex~kpc$^{-1}$ for O/H) from optical spectra of 6 \hii\
regions; {\em iii}) Magrini et al. (\cite{magrini07a}) who obtain an O/H gradient of
14 \hii\ regions, located in the radial range from $\sim$2 to
$\sim$7.2 kpc with a slope of $-0.054\pm0.011$~dex~kpc$^{-1}$; {\em iv})
Rosolowsky \& Simon (\cite{rs08}) who observe the largest sample of \hii\
regions, 61, finding a slope of $-0.027\pm0.012$~dex~kpc$^{-1}$ from $\sim$0.2 to
$\sim$6 kpc;
{\em v}) Rubin et al. (\cite{rubin08}) who obtain Ne/H and S/H gradients from 
Spitzer infrared spectra  of  -0.058$\pm$0.014~dex~kpc$^{-1}$  and -0.052$\pm$0.021~dex~kpc$^{-1}$, respectively.

 Rosolowsky \& Simon (\cite{rs08}) attribute the large
discrepancies in different authors' determinations to an intrinsic
scatter of about 0.1 dex around the average gradient, but  the scatter is unexplained by
the abundance uncertainties.  This kind of scatter is also seen in the
Milky Way  gradient from \hii\ regions (e.g. Afflerbach et al. \cite{aff97}), and it may be caused
by metallicity fluctuations in the ISM and by the spiral arms.  Thus a
limited number of observations, coupled with a significant metallicity
scatter at a given radius, may produce widely varying results. 
In the case of a shallow gradient this effect is even stronger; 
for example, for 
an abundance gradient of -0.02-0.03 dex kpc$^{-1}$ in a galaxy with a radius of 
10 kpc and a scatter of 0.1 dex, one would
need more than 30 \hii\ regions  to obtain  a good estimate of the slope of the gradient (Bresolin et al. \cite{bresolin09}).
Thus, only a large number of measurements can overcome the
uncertainties engendered by the intrinsic variance and relatively
shallow gradient of M33. 

Our sample is composed of 48 \hii\ regions.
We derived the physical 
and chemical properties  for 19 \hii\ regions  which have  not  been  observed previously, and 
for other 14 \hii\ regions whose chemical abundances have already been published.  
For the remaining \hii\ regions, the faintness of their spectra did not allow 
any reliable abundance determination, since their electron temperature (\te) could not be derived.
We complemented these observations by a sample of 102 planetary nebulae (PNe), 
already presented by Magrini et al. (\cite{magrini09}, hereafter M09), observed during the same run. 
The main  advantage  of  observing a combined  sample of \hii\ regions and PNe is being able 
to use not only the same observational set-up,  but also the same data reduction and analysis techniques,
and to use identical abundance determination methods.

Although our sample of  \hii\
regions does not add much to the literature, 
the presence of several objects in common with previous studies allows us to 
check the consistency  of different sets of chemical abundance results. 
By including at the same time two stellar populations of different 
ages but with similar spectroscopic characteristics, 
our observations allowed us to study for the first time the global metallicity, its 2-dimensional (2D)  distribution and its radial gradient,
at two different epochs in the galaxy's lifetime avoiding biases introduced by  different metallicity analysis.  
The aim of the present study is to settle the questions of the value of  the metallicity gradient in
M33 and its time evolution. In this framework, the new observations of \hii\ regions and PNe in M33 complemented 
with the previous data represent  the largest metallicity database available for an external galaxy.

In addition to metallicity data, recent results, such as the detection of inside-out 
growth in the disk of M33 (Williams et al. \cite{williams09}), the detailed analysis of the star formation  
both in the whole disk (Verley et al. \cite{verley09}) and in several giant \hii\ regions (Relano \& Kennicutt \cite{relano09}),  
stimulated us to revise the already existing chemical evolution model (M07b)
and the star formation process in M33. 
Particular attention was put on the observational constraints that our previous model failed to reproduce, such as the 
radial profile of the molecular gas and the relationship between the SFR and the molecular gas.

The paper is organized as follows. In Sect. \ref{sect_mmt} we describe 
our sample of \hii\ regions observed with MMT.
These data, together with a large literature dataset,  allowed us to
compute the metallicity gradient of \hii\ regions again.  In
Sect. \ref{sect_distr} we present the 2D distribution of the metallicity
and the radial gradient of different types of \hii\ regions.   
In Sect. 4 we discuss the off-centre metallicity peak and its 
origin. In Sect. \ref{sect_model} the data are compared 
with the prediction of chemical evolution model of M33.
Finally, our conclusions and a summary are given in
Sect. \ref{sect_conclu}.

\section{The \hii\ region dataset}
\label{sect_mmt}

Hot O-B stars ionize their surrounding medium,  producing the characteristic emission-line spectra of
\hii\ regions. 
The \hii\ regions of M33  studied in the literature span a wide range of luminosities. 
Their intrinsic brightness led  giant \hii\ regions to be preferred in the
earlier studies (e.g. Kwitter \& Aller \cite{kwitter81}, Vilchez et al. \cite{vilchez88}) when only
relatively small telescopes were available.  Smaller and fainter \hii\ regions  have instead been
the subject of later spectroscopic investigations
(e.g., Magrini et al. \cite{magrini07a}, Rosolowsky \& Simon \cite{rs08}).
The latest abundance determinations have been restricted to the \hii\ regions with available  electron temperature
measurements. 
Several emission-line diagnostics of  nebular \te\  
are indeed present in the optical spectrum of an \hii\ region, namely: \oiii\
4363 \AA, \nii\ 5755 \AA, \siii\ 6312 \AA, \oii\ 7320-7330 \AA.  Determining 
\te\ is the only way to derive  the ionic and total chemical abundances safely and
accurately. 
An assumed \te\ could produce error of a factor of 2 or more in the final
chemical abundances (cf., e.g., Osterbrock \& Ferland
\cite{ost06}).  This is why in the following analysis we include only those \hii\ regions whose \te\ is directly measured.

\subsection{The MMT observations: data reduction and analysis}

In November 2007, we obtained spectra of 48 \hii\ regions (and 102 PNe) in M33 using the 
MMT Hectospec fiber-fed spectrograph (Fabricant et al. \cite{fabricant05}).  The
spectrograph was equipped with an atmospheric dispersion corrector and
it was used with a single setup: 270 mm$^{-1}$ grating at a dispersion
of 1.2 \AA ~pixel$^{-1}$.  The resulting total spectral coverage
ranged from approximately 3600
\AA\ to 9100 \AA, thus including the basic emission-lines necessary
for determining  their
physical and chemical properties.  The instrument deploys 300 fibers
over a 1-degree diameter field of view, and the fiber diameter is $\sim$ 1.5\arcsec\  
(6 pc adopting a distance of 840 kpc to M~33).  

Some of the \hii\ regions in our sample already have published spectra in the literature so we use them as control sample,
while several are new.  In Table \ref{tab_pos} we list the \hii\ regions from the 
total observed sample for which we can derive the physical 
and chemical properties.
In Table \ref{tab_pos} we list the \hii\ regions from the 
total observed sample for which we can derive the physical 
and chemical properties. The identification names are
from: BCLMP-- Boulesteix et al. (\cite{boulesteix74});
CPSDP-- Courtes et al. (\cite{courtes87}); GDK99--Gordon et al. (\cite{gordon99}); EPR2003--Engargiola et al. (\cite{engargiola03}); MJ98--Massey \& Johnson (\cite{massey98}).  
The \hii\ regions not identified in
previous works are labled with
LGC-HII-n as in Magrini et al. (\cite{magrini07a}), standing for \hii\ regions discovered by the Local Group
Census project (cf. Corradi \& Magrini \cite{corradi06}).  The
coordinates J2000.0 of the position of the fibers projected on the sky are
shown in the third  and forth columns. They do not correspond exactly
to the centre of the emission line objects, but generally to the
maximum \oiii\ emissivity.

\begin{table}
\caption{MMT observations of \hii\ regions with chemical abundance determination. }
\label{tab_pos}
\scriptsize{
\begin{tabular}{llcc}
\hline\hline
New sample &ID     & RA        & Dec                 \\
    (1) &(2)     & (3)        & (4)                 \\
\hline
1 & BCLMP 275A   	& 1:32:29.5 & 30:36:07.90   \\
2 &GDK99 3     		& 1:32:31.7 & 30:35:27.39    \\
3&LGCHII14            & 1:32:33.3 & 30:32:01.90    \\
4&CPSDP 26    	     & 1:32:33.7 & 30:27:06.60   \\
5&LGCHII15          	& 1:32:40.8 & 30:24:24.99   \\
6&EPR2003 87  	& 1:32:42.4 & 30:22:25.59   \\
7&LGCHII16             & 1:32:43.5 & 30:35:17.29   \\
8&LGCHII17            & 1:33:11.3 & 30:39:03.39    \\
9&BCLMP 694   	& 1:33:52.1 & 30:47:15.59   \\
10&BCLMP 759   	& 1:33:56.8 & 30:22:16.50    \\
11&MJ98 WR 112 	& 1:33:57.3 & 30:35:11.09   \\
12&LGCHII18       	& 1:33:58.9 & 30:55:31.30   \\
13&BCLMP 282		&1:32:39.1 & 30:40:42.10 \\ 
14&BCLMP 264		&1:32:40.2 & 30:22:34.70 \\
15&BCLMP 238		&1:32:44.5 & 30:34:54.30 \\ 
16&BCLMP 239		&1:32:51.8& 30:33:05.20 \\ 
17&BCLMP 261		&1:32:54.1 & 30:23:18.70\\
18&CPSDP 123 	&1:33:20.4 & 30:32:49.20\\ 	
19&CPSDP 43A 	&1:33:23.9 & 30:26:15.00 \\ 
\hline
Control sample			&		  &			&\\
\hline
20 &LGCHII2             		     & 1:32:43.0 & 30:19:31.19     \\
21 &LGCHII3             		     & 1:32:45.9 & 30:41:35.50   \\
22 &BCLMP289            		& 1:32:58.5 & 30:44:28.60   \\
23 &BCLMP218            		& 1:33:00.3 & 30:30:47.30   \\
24 &MA1                 			& 1:33:03.4 & 30:11:18.70   \\
25 &BCLMP290            		& 1:33:11.4 & 30:45:15.09   \\
26 &IC132               			& 1:33:15.8 & 30:56:45.00   \\
27 &BCLMP45             		& 1:33:29.0 & 30:40:24.79   \\
28 &BCLMP670            		& 1:34:03.3 & 30:53:09.29   \\
29 &MA2                 			& 1:34:15.5 & 30:37:11.00   \\
30 &BCLMP691            		& 1:34:16.6 & 30:51:53.99   \\
31 &IC131					& 1:33:15.0 & 30:45:09.00\\	
32 & IC133					& 1:33:15.9 & 30:53:01.00\\ 
33 &BCLMP745				& 1:34:37.6 & 30:34:55.00 \\				 
\hline
\hline
\end{tabular}}
\end{table}

\begin{figure}
  \centering
   \includegraphics[angle=0,width=10cm]{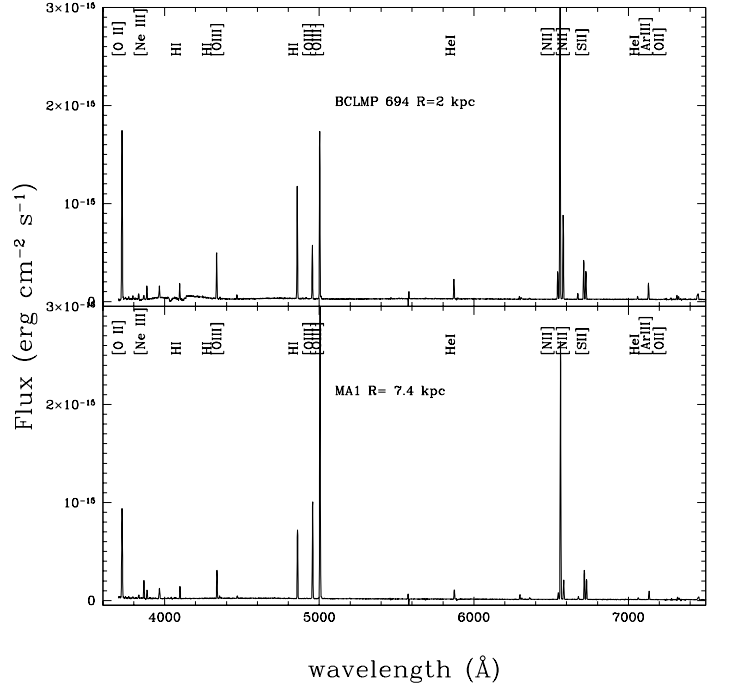}
    \caption{Two example spectra of \hii\ regions located at different galactocentric distance: BCLMP 694 at $\sim$ 2 kpc 
    and MA1 at $\sim$7.4 kpc.  }
             \label{Fig_sp}%
   \end{figure}

The spectra were reduced using the Hectospec package.  The relative
flux calibration was done observing the standard star Hiltm600 (Massey
et al.~\cite{massey88}) during the nights of October 15 and November 27.  The
emission-line fluxes were measured with the package SPLOT of
IRAF\footnote{IRAF is distributed by the National Optical Astronomy
Observatory, which is operated by the Association of Universities for
Research in Astronomy (AURA) under cooperative agreement with the
National Science Foundation}.  Errors in the fluxes were calculated
taking the statistical error in the measurement of the
fluxes into account, as well as systematic errors of the flux calibrations,
background determination, and sky subtraction.  The observed line
fluxes were corrected for the effect of the interstellar extinction
using the extinction law of Mathis~(\cite{mathis90}) with $R_V$=3.1.  We
derived \cbeta, the logarithmic nebular extinction, by using the
weighted average of the observed-to-theoretical Balmer ratios of
H$\alpha$, H$\gamma$, and H$\delta$ to H$\beta$ (Osterbrock \& Ferland
\cite{ost06}). The detailed description of the data reduction and the plasma and  chemical 
analysis can be found in Magrini et al. (\cite{magrini09}, hereafter M09).
Spectra of two \hii\ regions, one close to the galactic centre and one in the outer part of the disk 
are shown in Fig.\ref{Fig_sp}.

Table \ref{tab_flux} gives the results of our emission-line measurements and extinction
corrections for 33 \hii\ regions whose 
spectra were suitable for determining  physical and chemical properties.
The columns of Table \ref{tab_flux}
indicate:  (1) \hii\ region name; (2) nebular extinction coefficient
\cbeta\ with its error;  (3) emitting ion; (4) rest-frame wavelength 
in \AA;  (5) measured line fluxes; (6) absolute errors on the measured line fluxes; 
(7) extinction corrected line fluxes. Both F$_{\lambda}$ (5)
and I$_{\lambda}$ (7) are expressed on a scale where  \hb=100. Table
\ref{tab_flux} is published in its entirety in the  
electronic edition of \aap. A portion is 
shown here for guidance regarding its form and content.
The analysed \hii\ regions represent about 2/3 of our sample. The remaining 1/3 \hii\ regions have noisy spectra and 
are distributed at all galactocentric radii.
We used the extinction-corrected intensities to obtain the electron
densities and temperatures. Electronic density was derived from the intensities of the sulphur-line doublet
\sii\ 6716,6731 \AA.
We used the intensities of several emission-line ratios, when available,  to derive low and medium-excitation temperatures
(see also Osterbrock \& Ferland \cite{ost06}, $\S$5.2):
\nii $\lambda$5755/($\lambda$6548 + 
$\lambda$6584) and \oii  $\lambda$3727/($\lambda$7320 + 
$\lambda$7330) for low-excitation \te, while \oiii $\lambda$4363/($\lambda$5007 + $\lambda$4959) 
and  \siii $\lambda$6312/($\lambda$9069 + 
$\lambda$9532) for medium-excitation \te.   
We performed plasma diagnostics and ionic abundance calculation by using the 5-level atom model included in the {\it nebular}
analysis package in IRAF/STSDAS (Shaw \& Dufour~\cite{shaw94}). 
The elemental
abundances are then determined by applying the ionization correction
factors (ICFs) following the prescriptions by Kingsburgh \&
Barlow~(\cite{kb94}) for the case where only optical lines are
available. 
In the abundance analysis we adopted
\tenii\ and/or \teoii\ for computing  the N$^+$, O$^+$, S$^+$ abundances, while
\teoiii\ and/or \tesiii\  for  O$^{2+}$, S$^{2+}$, Ar$^{2+}$,
He${^+}$, and He$^{2+}$. 
We calculated the abundances of \hei\ and \heii\ using the
equations 
of Benjamin et al.~(\cite{benjamin99}) in two density regimes, i.e. $n_{e}$
$>$1000 cm$^{-3}$ and $\leq$1000 cm$^{-3}$.  The Clegg's collisional
populations were taken into account (Clegg \cite{clegg87}). 
In Table~\ref{tab_abu} we present the electron densities and temperature, and  the 
ionic and total chemical abundances of our \hii\ region sample, which only  includes  
\hii\ regions with at least one measured value of \te. The columns of Table~\ref{tab_abu}  present:  (1) identification name; 
(2) label of each plasma diagnostic and abundances available;  (3) 
relative values obtained from our analysis.
Table~3 is published  entirety in the 
electronic edition. 

We derived the temperature and density uncertainties
using the error propagation of the absolute
errors on the line fluxes. The errors on the ionic and total abundances were computed taking
the uncertainties in the observed fluxes, in the electron
temperatures and densities, and in \cbeta into account. 
In Table \ref{tab_tot_abu} a summary of the  total abundances   He/H, O/H, 
N/H, Ne/H, S/H. and Ar/H 
and their  errors are presented. 
The He abundance is shown by number with its absolute error, while the metal abundances 
are expressed in  the form of log(X/H)+12  with errors expressed in dex.  
The last row indicates the average abundances computed by number. 

\subsection{The PN data-set}

The PN population of M33 was studied by Magrini et al. (\cite{magrini09}) using multi-fiber
spectroscopy with Hectospec at the MMT with the same spectroscopic setup and during the same nights
as the \hii\ region observations presented here. 
Spectra of 102 PNe were analysed and  plasma diagnostics and chemical abundances
obtained for  93 PNe where the necessary diagnostic lines were measured.
The data reduction and the plasma diagnostics followed exactly the same procedure 
as described in the present paper, thus ensuring 
that no biases are introduced for the  different analysis of the spectra. 
About 20$\%$ of the studied PNe have young progenitors, the so-called Type I PNe.  
The rest of the PNe in the sample are the progenies of an old disk stellar population, with main
sequence masses M$<$3M${_\odot}$ and ages t$>$0.3~Gyr.
A tight relation between the O/H and Ne/H abundances was found, 
excluding that both elements have been altered by PN progenitors and supporting the validity of 
oxygen as a good tracer of the ISM composition at the epoch of the progenitors' birth.

\begin{table} 
\caption{Observed and de-reddened fluxes. }
\label{tab_flux}
\scriptsize{
\begin{tabular}{lllllll}
\hline\hline
ID & c({H$\beta$}) & Ion & $\lambda$ (\AA) & F$_{\lambda}$ & $\Delta$(F$_{\lambda}$) & I$_{\lambda}$ \\
(1)&(2)				&(3)	&(4)						&(5)				&(6)									&(7)\\
\hline
BCLMP275A &0.493$\pm$0.009     &[OII] &     3727  & 186.4   &    2.2   &       256.7\\    
&&[NeIII]/HI &3968&  3.2  &     0.7  &    	4.1\\      
&&HI        & 4100 &  13.0 &      1.1 &       16.2     \\   
&&HI        & 4340 &  34.4 &      1.1 &       39.8     \\   
&&[OIII]     &4363 &  1.1  &       0.8&      	1.2      \\     
&&HI         &4861 &  100.0&       1.6    &     100.1    \\
&&[OIII]     &4959 &  57.3 &      1.2     & 	55.8     \\ 
&&[OIII]     &5007 &  169.9   &    2.0    &  	163.3     \\ 
&&HeI        &5876 &  13.5    &   0.9     & 	10.7     \\ 
&&[SIII]     &6312 &  1.7     &  0.8      &	1.3     \\ 
&&[NII]      &6548 &  9.1     &  1.0      &	6.5     \\ 
&&HI         &6563 &  409.6   &   2.3     &   290.2  \\ 
&&[NII]      &6584 &  28.9    &   1.1     & 	20.4  \\    
&&HeI        &6678 &  3.9     &  0.8      &	2.7     \\ 
&&[SII]      &6717 &  32.7    &   1.1     & 	22.6      \\
&&[SII]     & 6731 &  23.1    &   1.2     & 	15.9      \\
&&HeI       & 7065 &  2.6     &  0.8      &	1.7      \\
&&[ArIII]    &7135 &  9.8     &  0.8      &	6.4      \\
&&[SIII]     &9069 &  21.4    &   1.0     & 	10.2 \\
     \hline
GDK99 3 &0.448$\pm$0.007 & [OII]     & 3727  & 83.3 &      1.0   &       111.4 \\
&&HI         &3835 &  6.8     &     0.6    &        8.8     \\  
&&[NeIII]    &3869 &  25.9    &   0.6      &	33.4      \\
&&HeI        &3889 &  9.8     &  0.6      &	12.6      \\
&&[NeIII]/HI &3968 &  18.0    &   0.7     & 	22.7      \\
&&HI         &4100 &  17.5    &  0.7      &  21.2   \\
&&HI         &4340 &  40.0    &  0.9      &  45.8   \\
&&[OIII]     &4363 &  3.3     &  0.5      &	3.8  \\    
&&HeI        &4471 &  3.5     &  0.5      &	3.8    \\  
&&HeII       &4686 &  0.5     &  0.5      &	0.6      \\
&&HI         &4861 &  100.0   &    1.3    &     100.1    \\
&&[OIII]     &4959 &  196.2   &    1.5    &  	191.5     \\ 
&&[OIII]     &5007 &  585.9   &    2.8    &  	565.0      \\
&&HeI        &5876 &  14.0    &   0.7     & 	11.3      \\
&&[NII]      &6548 &  6.1     &  0.4      &	4.5      \\
&&HI         &6563 &  396.4   &   1.9     &   289.9   \\
&&[NII]      &6584 &  19.2    &   0.6     & 	14.0   \\   
&&HeI        &6678 &  4.3     &  0.4      &	3.1      \\
&&[SII]      &6717 &  31.0    &   0.7     & 	22.2      \\
&&[SII]     & 6731 &  21.6    &   0.7     & 	15.4      \\
&&HeI       & 7065 &  3.4     &  0.5      &	2.3      \\
&&[ArIII]    &7135 &  14.8    &   0.5     & 	9.9      \\

\hline
\hline
\end{tabular}}
\end{table}

\begin{center}
\begin{table} 
\caption{Plasma diagnostics and abundances. }
\label{tab_abu}
\scriptsize{
\begin{tabular}{cll}
\hline\hline
ID & & \\
(1)&(2)&(3)\\
\hline
BCLMP725A & 						& \\
             &\teoiii    		&10600 \\ 
             & \tesiii          &15800\\
             &HeI/H               & 0.076\\
			& [OII]/H 			    &7.740e-05\\
	         & [OIII]/H              & 4.737e-05\\
             & ICF(O)  		    &1.000\\
             & O/H  			    &1.248e-04\\
             &[NII]/H  				&3.287e-06\\
             &ICF(N)              &1.612\\
             &N/H    				&5.299e-06\\
             &[ArIII]/H 			&4.910e-07\\
			&ICF(Ar) 			&1.87\\
			&Ar/H 				&9.182e-07\\
			&[SII]/H 				&7.486e-07\\
			&[SIII]/H 				&1.100e-06\\
			&ICF(S) 			&1.019\\
			&S/H 				&1.884e-06\\
\hline
GDK99 3& 				& \\
	          &\teoiii   & 10200 \\
			 & HeI/H 	& 0.081 \\
			 &[OII]/H 		& 3.880e-05\\
			 &[OIII]/H 	&1.860e-04\\
			 & ICF(O)  & 1.006 \\
			 & O/H  		&2.261e-04\\
			 &[NII]/H  		&2.536e-06\\
              &ICF(N)  	&5.826\\
			 & N/H   		& 1.475e-05\\
			 &[NeIII]/H 	&3.150e-05\\
			 &ICF(Ne)  &1.215\\
			 &Ne/H       &3.829e-05\\
			 &[ArIII]/H 	&8.560e-07\\
			 &ICF(Ar) 	&1.87\\
			 &Ar/H 		&1.601e-06\\
			 &[SII]/H 		&8.157e-07\\
			 &ICF(S) 	&1.323\\
			 & S/H 		&8.255e-06\\
\hline
\hline
\end{tabular}
}
\end{table}
\end{center}

\begin{center}
\begin{table*}
\caption{The chemical abundances of our MMT sample.  }
\label{tab_tot_abu}
\scriptsize{
\begin{tabular}{lllllll}
\hline\hline
\#    				& He/H        & O/H 		& N/H 	& Ne/H 	& Ar/H 	& S/H                \\
(1)					&(2)			&(3)	     &(4)		&(5)		&(6)		&(7)\\
\hline
1         & 0.076$\pm$0.001	    &8.10$\pm$0.07          & 6.72$\pm$0.15     	&   -  					& 5.96$\pm$0.34     	& 6.27$\pm$0.20\\
2         & 0.081$\pm$0.001       & 8.35$\pm$0.05		& 7.17$\pm$0.11     	&  7.58$\pm$0.015& 6.20$\pm$0.09    & 6.92$\pm$0.21 \\
3         & 0.058$\pm$0.003       &7.97$\pm$0.08		& 6.88$\pm$0.12		&-		      			& 6.02$\pm$0.11    & 6.62$\pm$0.15\\
4         & 0.088$\pm$0.010       &8.36$\pm$0.05		& 7.24$\pm$0.19		&-		      			& 5.82$\pm$0.18    & 6.46$\pm$0.19\\
5         & 0.085$\pm$0.005       &8.45$\pm$0.02          & 7.35$\pm$0.14     	&-					   	&-				& 6.59$\pm$0.12\\
6         &0.114$\pm$0.003        &8.52$\pm$0.08			& 6.86$\pm$0.15     	&7.60$\pm$0.20     	& 6.27$\pm$0.12   & 6.50$\pm$0.20\\
7         & 0.085$\pm$0.005       &8.30$\pm$0.12          & 7.09$\pm$0.20     	&7.00$\pm$0.18     	& 6.07$\pm$0.16   & 6.61$\pm$0.15\\
8         & 0.113$\pm$0.005       &8.21$\pm$0.07			& 7.30$\pm$0.24      	&7.69$\pm$0.42    	& 6.20$\pm$0.20   & 6.49$\pm$0.15\\
9         &0.092$\pm$0.008        &8.33$\pm$0.10		& 7.21$\pm$0.10		& 7.53$\pm$0.20  		&6.22$\pm$0.30	&   6.79$\pm$0.10\\
10		 & 0.077$\pm$0.005       &8.11$\pm$0.08			& 6.88$\pm$0.29      	& -   					& 6.16$\pm$0.22	&   6.66$\pm$0.15\\
11        & 0.093$\pm$0.003       &8.30$\pm$0.07		& 7.37$\pm$0.02	 	& -							&  6.30$\pm$0.30	&   7.09$\pm$0.11\\
12		 & 0.091$\pm$0.008	     &8.15$\pm$0.07		&7.94$\pm$0.05		&7.20$\pm$0.10			&6.24$\pm$0.10			&6.51$\pm$0.20\\
13	       &0.080$\pm$0.005           & 8.32$\pm$0.05		&  6.92$\pm$0.12		&-						&6.19$\pm$0.12       &6.36$\pm$0.18   \\
14        &0.076$\pm$0.003            &7.87$\pm$0.08   		& 6.93$\pm$0.15		&   7.11$\pm$0.20		& 5.89$\pm$0.20		& 6.41$\pm$0.18\\   
15		 &0.089$\pm$0.006		    &7.92$\pm$0.08  		& 6.75$\pm$0.15		&   -   						& 5.92$\pm$0.18		& 6.57$\pm$0.15\\   
16		 &0.086$\pm$	0.005		    &8.27$\pm$0.07		& 7.20$\pm$0.14		&   -						&  -   					& 6.52$\pm$0.20\\   
17        &0.115$\pm$0.008		    &8.01$\pm$0.08		& 6.62$\pm$0.16		&  7.16$\pm$0.18		& 5.93$\pm$0.14		& 6.40$\pm$0.18\\   
18        &0.078$\pm$0.005		    &8.35$\pm$0.05		& 7.35$\pm$0.12		&   -   						& 6.35$\pm$0.15		& 6.94$\pm$0.20\\  
19		 &0.059$\pm$0.003		    &8.40$\pm$0.05         & 8.05$\pm$0.10		&   8.03$\pm$	0.18		& 5.16$\pm$0.18		&  7.02$\pm$0.17\\   
20   	 &0.083$\pm$0.001	    & 8.08$\pm$0.05		& 6.91$\pm$0.15	 	&   6.87$\pm$0.18	&  6.18$\pm$0.25	&   6.64$\pm$0.15\\
21    	 &0.086$\pm$0.001	    &8.42$\pm$0.06		& 7.56$\pm$0.13	 	&   7.31$\pm$0.15	&  6.01$\pm$0.27	&   6.63$\pm$0.10\\
22        & 0.072$\pm$0.005       & 8.35$\pm$0.12		& 7.34$\pm$0.16	 	&   7.75$\pm$0.38 &  5.67$\pm$0.50   &   6.89$\pm$0.11\\
23       &0.088$\pm$0.001        & 8.17$\pm$0.12		& 6.97$\pm$0.18	 	&   -					&   6.15$\pm$0.14  &  6.73$\pm$0.23 \\
24        &0.080$\pm$0.008        & 8.28$\pm$0.15		& 7.10$\pm$0.20	 	&   7.61$\pm$0.28	&   6.14$\pm$0.32  & 6.71$\pm$0.40\\
25        &0.096$\pm$0.005	    &8.38$\pm$0.13		& 7.37$\pm$0.15	 	&   7.36$\pm$0.20	&   5.84$\pm$0.10	& 6.57$\pm$0.15\\
26 		 &0.061$\pm$0.003	    &  7.98$\pm$0.05		&6.98$\pm$0.15		&   7.28$\pm$0.12	&   5.86$\pm$0.15	& 6.36$\pm$0.13\\
27        &0.095$\pm$0.001	    &  8.48$\pm$0.08		&  7.62$\pm$0.12	 	&   7.73$\pm$0.15 &  6.33$\pm$0.25	&   6.87$\pm$0.15\\
28        &0.088$\pm$0.002        & 8.30$\pm$0.07		&7.10$\pm$0.18		&  7.42$\pm$0.20	&6.23$\pm$0.30	&   6.63$\pm$0.20\\
29        & 0.091$\pm$0.005		& 8.31$\pm$0.10		&  7.19$\pm$0.15     	&   7.46$\pm$0.25	 &   6.42$\pm$0.24  &   6.92$\pm$0.21\\
30        &0.096$\pm$0.003		& 8.42$\pm$0.06		&  7.20$\pm$0.12	& 7.85$\pm$0.21	 	&   6.37$\pm$0.32  & 6.75$\pm$0.23\\
31			&  0.097$\pm$0.005		&8.47$\pm$0.08			&  7.26$\pm$0.15	&   7.76$\pm$0.20	&   6.29$\pm$0.25		& 7.06$\pm$0.15\\   
32			& 0.079$\pm$0.005		&8.27$\pm$0.08			&  7.21$\pm$0.17	&   7.58$\pm$	0.21	&   5.49$\pm$	0.30		&  6.83$\pm$0.15\\   
33			&0.067$\pm$0.008		&7.93$\pm$0.10			&  7.10$\pm$0.20	&   -		 			&  6.09$\pm$0.21		&   6.51$\pm$0.20\\   
\hline
&						&							&						&						&							&						\\
				&0.085$\pm$0.011	&8.27$_{-0.17}^{+0.12}$ & 7.31$_{-0.35}^{+0.30}$&	7.56$_{-0.30}^{+0.18}$	&6.13$_{-0.22}^{+0.14}$ & 	6.71$_{-0.34}^{+0.19}$ \\					&							&						\\
\hline
\end{tabular}}
\end{table*}
\end{center}

\section{The metallicity distribution in M33}
\label{sect_distr}

The large amount of chemical abundance data from \hii\ regions in M33 allow 
us to analyse the spatially-resolved distribution of metals in the ISM.  
In this section, we present the
radial distribution and the map of O/H, 
using the new data presented in this paper and all previous oxygen determinations for which
\te\ has been measured.  

\subsection{The metallicity gradient of \hii\ regions}
\label{sect_grad}

\begin{figure}
  \centering
   \includegraphics[angle=-90,width=8cm]{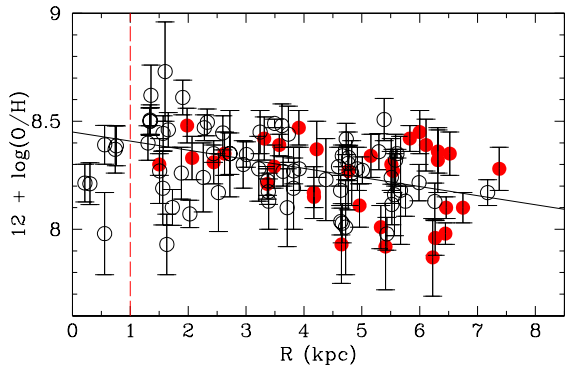}
    \caption{The O/H radial gradient for the cumulative \hii\ region sample: filled circles are the MMT 
    observations (new and control samples), empty circles are the literature abundances. 
    The continuous line is the  weighted linear least-square fit of Equation 3, i.e., with a radial range 
    from 1 to 8 kpc from the M33 centre. The dashed vertical line indicates the regions located at its left-side and 
    excluded from the fit. }
             \label{Fig_oxy}%
\end{figure}

Our cumulative sample includes: {\em i)} \hii\
regions by Magrini et al. (\cite{magrini07a}), which includes abundances from their own sample  
and  previous abundance determinations (all with \te\, and with abundances recomputed uniformly); 
{\em ii)} the sample by Rosolowsky \&
Simon (\cite{rs08}); {\em iii)} the present sample (see Table \ref{tab_abu}).  
In Fig.~\ref{Fig_oxy}, we show the oxygen abundance as a function of
galactocentric distance for the cumulative sample of \hii\ regions. 
In this figure each point corresponds to a single region; i.e., we do not plot multiple measurements 
for the same region but only the value with the lowest error. 
Note the large dispersion in the radial region between 1 and 2 kpc from the centre caused by several high- and low-metallicity 
regions, located in the southern arm (see Sect. 3.3), which might be related to the presence of a bar  
(e. g., Corbelli \& Walterbos \cite{corbelli07}).
We applied the routine
{\it fitexy} in Numerical Recipes (Press et 
al. 1992) to fit the relation between the oxygen abundances and the galactocentric distances, 
taking their errors into account and minimizing $\chi^2$.
Typical errors on the de-projected galactocentric distances associated 
with the uncertainty on the inclination were less than 0.1 kpc (Magrini et al. \cite{magrini07a}). 
The fit to the complete sample gives a gradient of
\begin{equation}
12 + {\rm log(O/H)} = -0.037 (\pm 0.009) ~  {\rm R_{GC}} + 8.465 (\pm
0.038) 
\label{eq1}
\end{equation}
where R$_{\rm GC}$ is the de-projected galactocentric distance in kpc,
computed by assuming an inclination of 53$^\circ$ and a position angle
of 22$^\circ$. 
A weighted linear least-square fit to the MMT sample only gives a gradient of
\begin{equation}
12 + {\rm log(O/H)} = -0.044 (\pm 0.017) ~  {\rm R_{GC}} + 8.447 (\pm0.084),
\label{eq2}
\end{equation}
which is  consistent within the errors with the gradient from the larger sample.
In the rest of the paper, we use the larger and more complete sample  when discussing the metallicity
gradient and its possible time variation,  but excluding the first kpc region 
where only a few low metallicity regions were analysed. We  discuss
the possible reasons for the lower value of the  central metallicity  later in this section and in Sect. \ref{sec_centre}.
The O/H gradient of the whole sample of \hii\ regions sample, excluding the central 1 kpc, is 
\begin{equation}
12 + {\rm log(O/H)} = -0.044 (\pm 0.009) ~  {\rm R_{GC}} + 8.498 (\pm0.041).
\label{eq3}
\end{equation}

\begin{figure}
  \centering
  \includegraphics[angle=-90,width=8cm]{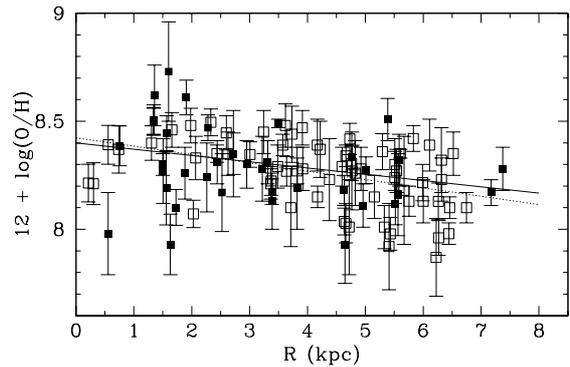}    
  \includegraphics[angle=-90,width=8cm]{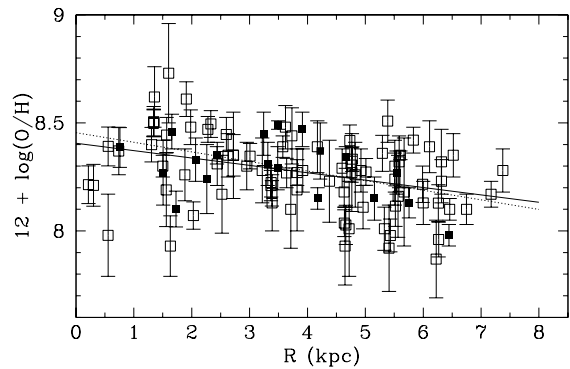}  
   \caption{The O/H radial gradient obtained in different regions of M33. Top: nearest side (filled squares) and 
    farthest side (empty squares) gradient. Bottom: North (filled squares) and South (empty squares) gradient.
    In each panel the weighted linear least-square fits of the two regions are shown with 
    two lines (continuous and dotted). }
             \label{Fig_oxy_nswe}%
              \end{figure}

We also  checked  the metallicity gradient 
by tracing it in different 
areas of the galaxy, namely in the northwest and in the southeast halves,  separately,
and in the nearest  and in the farthest sides. 
The results are shown in Fig. \ref{Fig_oxy_nswe}. The northern and southern gradients, as 
well as those relative to the nearest and farthest sides, are identical within the uncertainties, 
with slopes around -0.03-4 dex kpc$^{-1}$.  
The only difference found between the metallicity gradients obtained for sections of the galaxy is the presence 
of a high metallicity peak in the southern arm.
In conclusion, the present  \hii\ region sample (literature $+$ present-work, 103 objects), 
including only nebulae with measured \te, reinforce the recent results on the slope of 
the M33 O/H gradient, with a global slope up to around 8 kpc of -0.03 dex kpc$^{-1}$, 
and of about -0.04 dex kpc$^{-1}$ excluding the central 1 kpc. 
The very central regions remain somewhat undersampled (6 objects within 1 kpc from the 
centre, and 9 within 1.5 kpc) and in disagreement with 
other  results. 
A comparison with the metallicity gradient derived from young stars, which 
are representative of the same epoch in the lifetime of the galaxy as \hii\ regions, and with 
the infrared spectroscopy of \hii\ regions, is necessary.

Stellar abundances were obtained by Herrero et al.~(\cite{herrero94}) for
AB-supergiants, McCarthy et al.~(\cite{mccarthy95}) and Venn et
al.~(\cite{venn98}) for A-type supergiant stars, and Monteverde et
al. (\cite{monteverde97}, \cite{monteverde00}) and Urbaneja et
al.~(\cite{urbaneja05}) for B-type supergiant stars.  The largest sample of Urbaneja et al.~(\cite{urbaneja05}) 
gave a O/H gradient of -0.06$\pm$0.02~dex~kpc$^{-1}$.  
Recently, U et al. (\cite{U09}) has presented spectroscopic observations of a set of  A
and B supergiants. They determined  stellar metallicities
and derived the metallicity gradient in the disk of M33, finding solar metallicity
at the centre and 0.3 solar in the outskirts at a distance of 8 kpc.
Their average  metallicity gradient is -0.07$\pm$0.01 dex kpc$^{-1}$.
At a given radius, \hii\ regions  have abundances slightly
below the stellar results, and this is  probably due to the depletion of oxygen in \hii\ regions on dust grains
(e.g., Bresolin et al. \cite{bresolin09}).  
The slopes of the two gradients  agree if the comparison is done between about 1 ant 8 kpc.
The cause of  the difference between the supergiant and \hii\ region gradient  is  the metallicity value in the central regions. 
In fact, the \hii\ regions located within 1 kpc from centre   have metallicity below solar, whereas 
the supergiants are metal rich,  ranging from solar values to above solar. 
The origin of this discrepancy is not the 
temperature gradients within the nebulae (Stasi{\'n}ska 2005) because they become important at higher metal abundances. 

Recent observations of 25 \hii\ regions by Rubin et al. (\cite{rubin08}) with {\em Spitzer}
have allowed a measurement of the Ne/H and S/H gradient across the disk of M33 showing no 
decrease in chemical abundances in the central regions. Infrared Ne and S emission lines  do not have a strong dependence 
on \te, and consequently their abundances can be determined  even without a temperature measurement. 

One way to explain the low metallicity in the central 1.0$\times$1.0 kpc$^{2}$ 
area is related to the criterion used for  \hii\ regions.
Usually, chemical abundances derived from optical spectroscopy rely on direct measurement of the electron temperature, given by the 
\oiii\ 4363\AA\, emission line. This emission line is inversely proportional to the oxygen abundance
and barely detectable for O/H$>$8.6 from an average luminosity nebula (e.g., Nagao et al. \cite{nagao06}). 
The request for \oiii\ 4363 \AA\ detection might determine a bias towards lower metallicity, 
with the exclusion of the highest metallicity nebulae.
This could explain  the differences between the optical spectroscopy results
and both the stellar abundance determinations, and the \hii\ region infrared spectroscopy. 

\subsection{The abundance gradients of the other chemical elements}
\label{sec_other}

Our analysis allowed us to measure other chemical elements in addition to oxygen,  as He/H, 
N/H, Ne/H, S/H. and Ar/H.
The more reliable measurement is that of oxygen for the reasons illustrated in the Appendix,
 and we use it to follow the chemical evolution of M33. 
Nevertheless the other chemical elements are measured in enough  \hii\ regions    
to compute their radial gradients. 

In Table \ref{tab_other} we show the slopes and central abundance values of the radial gradients of 
N/H, Ne/H, S/H. and Ar/H of our sample of \hii\ regions. 
We did not calculate the radial gradient of He/H because we measured only the ionized fraction 
of He in \hii\ regions, which is only a small part of the total helium abundance.  
All gradients have a negative slope, consistent, within the errors, with the slope found for O/H, while 
N/H is a bit steeper, as already noticed, e.g., by Magrini et al.(\cite{magrini07a}). Its different  behaviour with 
respect to the $\alpha$-elements, as oxygen, neon, sulphur, and argon, comes from 
 the different places of production. $\alpha$-elements are indeed produced
by SNe II, which are the final phase of the evolution of massive stars, while nitrogen is one of the final products
of the evolution of long-lived low- and intermediate-mass stars. This is discussed in detail in Sec.\ref{sect_model}.  
Finally, there is a  very good agreement of the S/H and Ne/H gradients with those derived 
from the infrared spectra of \hii\ regions by Rubin et al.
(\cite{rubin08}), for which  they found  a gradient of   
-0.058$\pm$0.014~dex~kpc$^{-1}$  for
Ne/H and -0.052$\pm$0.021~dex~kpc$^{-1}$ for S/H. 

\begin{table} 
\caption{The radial gradients of  N/H, Ne/H, S/H. and Ar/H }
\label{tab_other}
\begin{tabular}{lll}
\hline\hline

12 + log(X/H)               				&slope             		& central value      \\
(1)						&(2)				&(3)			     \\					
\hline
N/H      		          & -0.08$\pm$0.03				& 7.53$\pm$0.15			\\
Ne/H				    & -0.05$\pm$0.04				& 7.71$\pm$0.21				\\	
S/H				    & -0.06$\pm$0.02				& 6.41$\pm$0.11				\\	
Ar/H	  			  & -0.07$\pm$0.03				& 6.98$\pm$0.13				\\	
\hline
\hline
\end{tabular}
\end{table}

\subsection{The  population-dependent metallicity gradient: giant vs. faint and compact \hii\ regions}

We now examine whether  any selection effect can be responsible for 
the difference between the steep gradient found in the early studies 
and the shallower  gradient of this work.
To this goal, we subdivided the sample of \hii\ regions according to their 
projected size and  surface brightness in the \ha\ emission-line.  
Then, we computed the intrinsic luminosity and the radius in an
\ha\ emission-line calibrated map (courtesy of R. Walterbos) for each nebula of the whole sample.
Defining the  surface
brightness (SB) as the ratio between the total flux and the area expressed in arcsec$^2$, we
subdivided the sample according to their size and SB. 
Considering their size, we defined them as {\em small} if their  radius $R<$15\arcsec\ (60 pc at the distance 840 kpc) and 
as {\em large}    if $R\geq$15\arcsec. 
Considering their surface brightness, we define them as {\em bright} if their 
surface brightness SB$>$ 5.5$\times$10$^{-19}$ erg cm$^{-2}$ s$^{-1}$ arcsec$^{-1}$, 
and as {\em faint} if  SB is lower than this limit. 
The four combinations are allowed, i.e. \hii\ regions can be  {\em small} and either {\em bright} 
or {\em faint}, or {\em large} and again {\em bright} 
or {\em faint}. 
In Table  \ref{tab_sb} we show the galactocentric distance,  R$_{GC}$, the 
\ha\ observed total flux, F$_{H\alpha}$,  the  radius, $R$, and 
the SB, of the so-called {\em giant} regions, i.e. those with SB$>$ 5.5$\times$10$^{-19}$ erg cm$^{-2}$ s$^{-1}$ arcsec$^{-1}$ and $R\geq$15\arcsec.

\begin{figure}
  \centering
  \includegraphics[angle=-90,width=8cm]{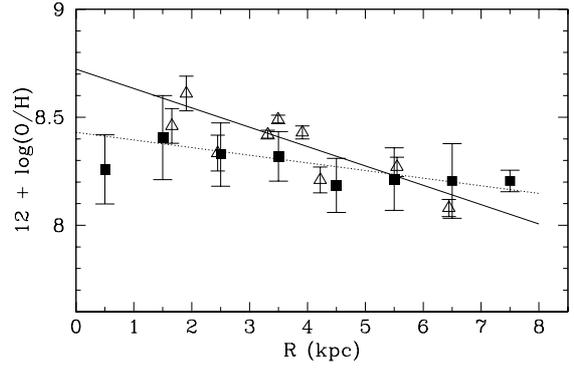}  
    \caption{The O/H radial gradient:  the giant \hii\ regions (empty triangles) 
    and the complete sample averaged in bins, each 1 kpc wide (filled squares). The continuous line is the 
    weighted  mean least square fit of the giant \hii\ region sample, while the dotted line refers to 
    the complete sample.  }
             \label{Fig_oxy_sb}%
   \end{figure}

To compare the total population with the {\em giant} \hii\ regions, we show in Fig. \ref{Fig_oxy_sb} 
the oxygen abundances of the cumulative sample, averaged in bins of  1 kpc each, 
together with the abundances of each single {\em giant} \hii\ region.
The O/H gradient of the \hii\ regions in Table \ref{tab_sb}, computed with a 
weighted linear least-square fit, is 
\begin{equation}
12 + {\rm log(O/H)} = -0.089 (\pm 0.023) ~  {\rm R_{GC}} + 8.72 (\pm0.09).
\label{eq3}
\end{equation}
The gradient of the remaining sample is the same of given in Eq.\ref{eq1}.
The {\em giant} regions show a significantly steeper gradient, consistent
with the gradients by Smith~(\cite{smith75}), Kwitter \& Aller~(\cite{kwitter81}), 
V\'{\i}lchez et al.~(\cite{vilchez88}), and Garnett et al.~(\cite{garnett97}). 
The question is whether this gradient is really significantly different from  
the whole sample,  
and if this is the case, what is the reason of such different behaviour.

Owing to the small number of {\em giant} \hii\ regions (9 in total), the uncertainty on the slope of their gradient
is high. Thus it could  still be in partial agreement, within the errors, with the larger sample, and
their difference might stem from metallicity fluctuations in the ISM.
On the other hand, the characteristics of the {\em giant} regions might be truly different from the average sample.   
For example large self-bound units are not destroyed by massive stars and thus retain their original 
structure and get continuously enriched by SF. 
However, while this might be the case for the metallicity peak near the centre, it does 
predict that giant regions should have higher metallicity at all galactocentric radii, which is clearly not the case. 
Moreover, in a recent paper, Rela{\~n}o \& Kennicutt (\cite{relano09}) studied the star formation in 
luminous \hii\ regions in M33, which correspond mostly to our {\em giant} \hii\ regions. 
They found that the observed UV and \ha\ luminosities are consistent 
with a young stellar population (3-4 Myr), born in an instantaneous burst.
Thus the steeper gradient might result form a combination of a small statistics and of 
a metal self-enrichment effect in the  {\em giant} region sample. 


\begin{table} 
 \caption{{\em Giant} \hii\ regions with derived chemical abundances.}
\label{tab_sb}
\scriptsize{
\begin{tabular}{lllll}
\hline
\hline
Id    & R$_{GC}$       & F$_{H\alpha}$                         &   R              & SB \\
              & (kpc)               &(10$^{-15}$erg/cm$^2$s)         & (arcsec)      &(10$^{-19}$erg/cm$^2$s arcsec$^2$)            \\
\hline
NGC595       	&	1.7   &       13.0 	 &        25 &  16.1 \\
C001Ab       	&	1.9   &  	    3.2   	&       20 &  6.2 \\
MA2       		&	2.5   &     	4.5   	 &      25 &  5.6 \\
BCLMP691    &   3.3   &    		2.4      &      15 &  8.2 \\
NGC604       	&	3.5   &   		36.6    &      30 &  31.4  \\
IC131          	&	3.9    & 		1.9    &      15 &  6.6  \\
BCLMP290     &   4.2    &     	2.2    &      15 &   7.5  \\
NGC588       	&	5.5    &    	4.2     &     20 &  8.0 \\
IC132       	     &	6.4    &    		3.5     &     20 &  6.8 \\
\hline
\hline
\end{tabular}
}
\end{table}

\subsection{The 2-dimensional distribution of metals}
\label{sect2d}
\begin{figure}
  \centering
  \includegraphics[angle=90,width=13cm]{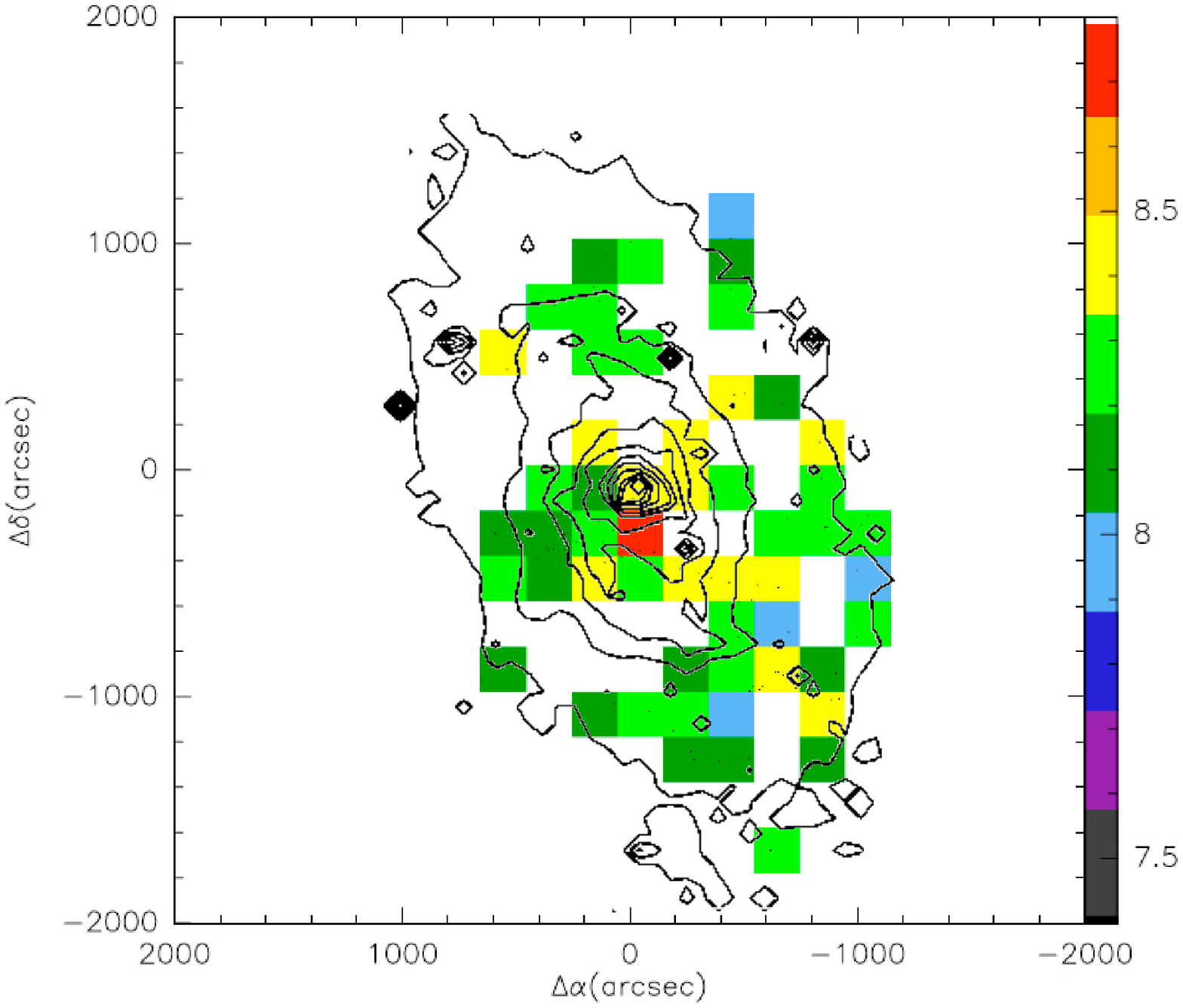}  
  \includegraphics[angle=90,width=13cm]{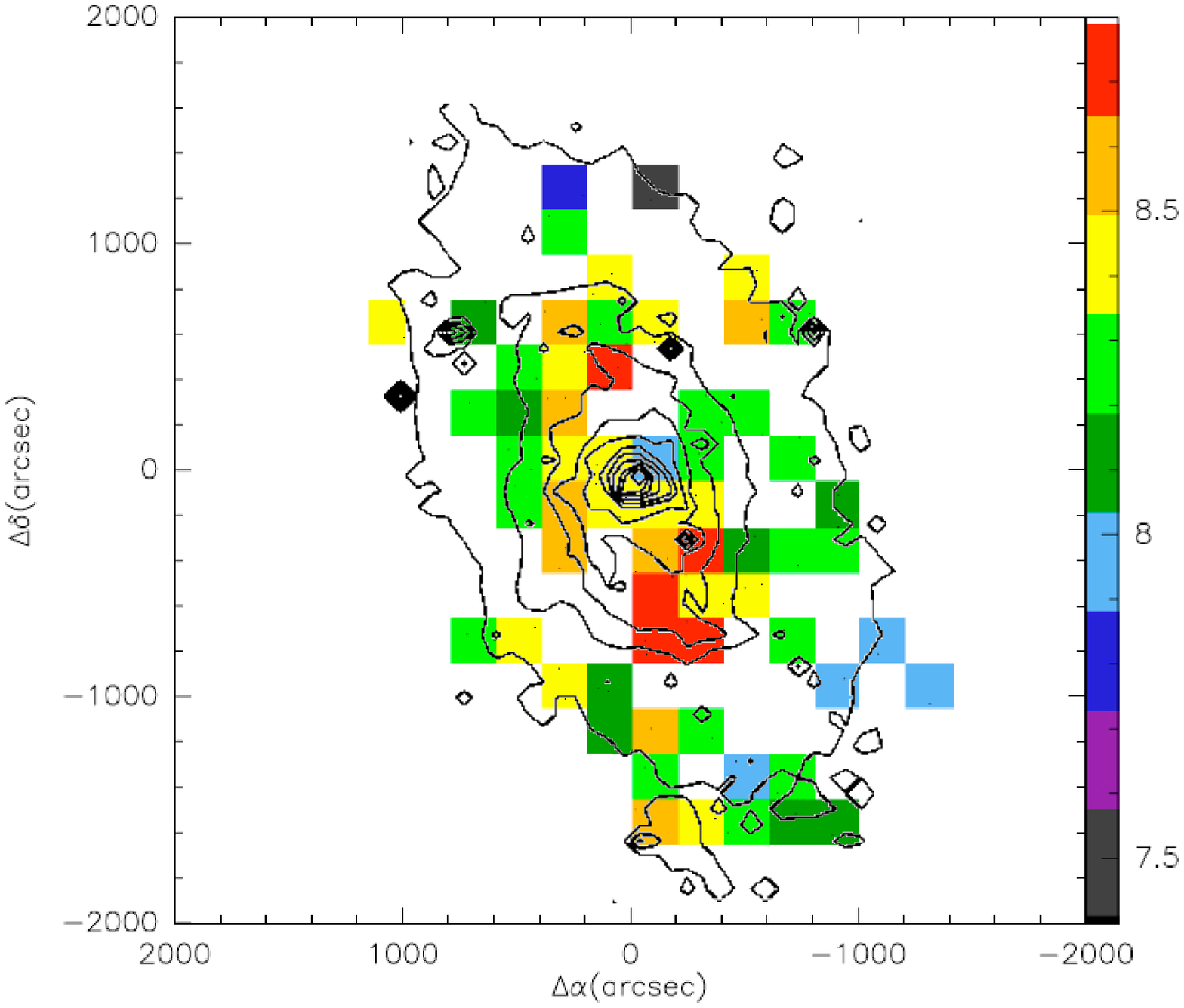}
    \caption{The oxygen abundance maps of M33 (60\arcmin $\times$ 60\arcmin): \hii\ regions (top) and PNe (bottom).   
    The O/H abundances are averaged in bins of 0.8$\times$0.8 kpc$^2$. 
    The colour-scale shows the oxygen abundance as indicated in the label. North at the top, east to the left. 
     The optical centre of M33 is located at (0,0). 
     The contour levels represent the stellar mass distribution derived from the JHK image of M33.}
             \label{fig_2dmap}%
   \end{figure}

The usual way to study the metallicity distribution in disk galaxies
is to average it azimuthally, assuming that {\em i}) the centre of the
galaxy coincides with the peak of the metallicity distribution and  {\em ii}) the metallicity 
distribution is axially symmetric. 
The large number of metallicity measurements in M33, both
from \hii\ regions and from PNe, allowed us to reconstruct not only
their radial gradient, but also their spatial distribution projected
onto the disk.  
In Fig. \ref{fig_2dmap}, we show the two-dimensional metallicity
distributions for M33 from \hii\ regions and from PNe  superimposed to a contour map of the stellar 
mass distribution derived from the JHK image, a composition of the image of Regan \& Vogel (\cite{regan94}) 
and the 2MASS image.
The O/H abundances were averaged in bins of 0.8$\times$0.8 kpc$^2$. 
The white pixels indicate  areas where metallicity measurements 
are lacking: for \hii\ regions they generally correspond to the interarm regions, while 
for PNe to spiral arms. 
 
The \hii\ regions with the highest metallicity are not located in the optical centre of the
galaxy (0,0 in the map), but rather lie at  radius 1-2 kpc in the southeast direction.  
Also in the case of PNe,
most of the metal rich PNe are located in the southern part of M33, from 2 to 4 kpc from the centre. 
However the lack of known PNe 
in the northern spiral arm at the same distance of the southern metal-rich PNe 
(because of the extended \hii\ regions not allowing the identification  of stellar emission-line sources)
does not allow a complete 2D picture of their metal distribution
around the central regions. 

\begin{figure}
  \centering
  \includegraphics[angle=-90,width=8cm]{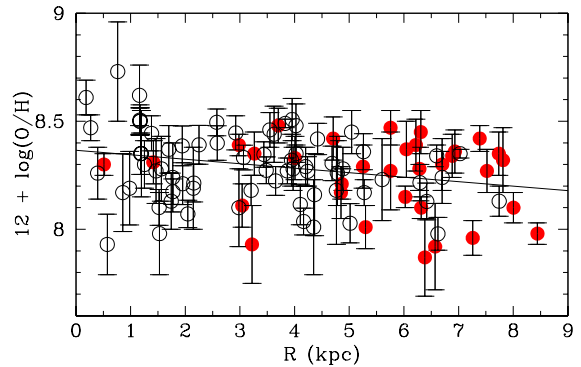}  
    \caption{The O/H radial gradient computed with the centre located at  RA 1:33:59  dec 30:33:35 (J2000.0)  to minimize the 
dispersion  in the slope of the radial gradient. }
             \label{Fig_oxy_off}%
   \end{figure}

To estimate the location of the off-centre metallicity peak for the \hii\ region map, we 
divided its squared 10\arcmin$\times$10\arcmin\ region, centered at RA 1:33:50.9 dec 30:39:36
(M33 centre from the 2MASS survey, Skrutskie et al. \cite{2mass}),  
with a 10$\times$10 grid.  
We computed the radial O/H gradients varying the central position in the grid and then  finding the one 
that minimizes the scatter of the gradient. 
We found an off-centre position at RA 1:33:59  dec 30:33:35 (J2000.0), which corresponds to the location 
of the high-metallicity \hii\ regions in the southern arm. 
The oxygen gradient measured from this central position of the whole \hii\ region population, including also the 
central objects, is   
\begin{equation}
12 + {\rm log(O/H)} = -0.021 (\pm 0.007) ~  {\rm R_{GC}} + 8.36 (\pm0.03), 
\label{eq5}
\end{equation}
and it is shown in Fig. \ref{Fig_oxy_off}.
The gradient of Eq. \ref{eq5} is flatter and with a somewhat lower absolute dispersion than the one in Eq.\ref{eq1}.
The reduction of the dispersion in the O/H gradient due to the 
displacement of the galaxy centre  is not enough strong  to confirm that 
the metallicity maximum corresponds to the real centre of the metallicity distribution.

\section{Why is  the metallicity peak off-centre? }
\label{sec_centre}
In the following, we examine several possibilities to explain
the presence of the off-centre metallicity maximum and the low
metallicity measured in the central region:
{\em i}) a local effect of ISM metallicity fluctuations; {\em  
ii})  the lack of  dominant gravitational centre
in the galaxy; {\em iii}) the selection criterion of \hii 
\ regions for metallicity determination.
\subsection{Local metallicity fluctuations}
Simon \& Rosolowsky (\cite{rs08}) have already noticed the  
non-axisymmetric distribution  of \hii\ region abundances and
suggest that the material enriched by the most recent
generation of star formation in the arm might not have been azimuthally
mixed through the galaxy. A strong OB association located in the  
southern arm might be responsible for the enhancement enrichment at
the location of the metallicity peak. Velocity shear is present in 
M33 even at small radii because of the
slow rise in  the rotation curve (Corbelli et. \cite{corbelli03}). 
At the peak location, differential rotation  will shear up the bubble of 
metals produced in the surrounding ISM in about 10$^8$~yrs. The
timescale seems  long enough to allow vigorous star formation
at a particular location to enrich the ISM of metals  well above the 
average value. However, the large dispersion in the metallicity
around the peak location seems to rule out an inefficient
azimuthal mixing or redistribution of the metals.

\subsection{The lack of a gravitational centre}
The non-axysymmetric metallicity distribution might  
be related
to a general non-axisymmetric character of central regions of M33, 
noticed in the past by  several authors.
Colin \& Athanassoula (\cite{colin81}) 
found that  the young population displacement is located
towards the southern side of M33 and amounts to approximately
2-3 \arcmin, i.e. 480-720 kpc.
Using evidence of other asymmetries in the  
inner regions of M33, such as those present in
the distribution of \hi\ atomic gas, of \hii\ regions, and in the 
kinematics,  
they proposed a bulge centre presently located in the  
northern part of the galaxy, which
is rotating retrogradely around the barycenter of the galaxy.
The analysis of infrared images (Minniti et al. \cite{minniti93}), however,
seems to point out to a small bulge with a much smaller displacement
the one advocated by Colin \& Athanassoula (\cite{colin81}).
A detailed analysis of the kinematics of the innermost regions of M33 
by Corbelli \&  Walterbos
(\cite{corbelli07}) confirms asymmetries in the 
stellar and gas  velocities, which however seem more related to
the presence of a weak bar. The exact galaxy centre is uncertain  on scales of a few arcsec. 
Thus, even if M33 lacks of a dominant gravitational centre
of M33 and the bright central cluster might migrate around it,
it seems unlikely that  the centre of the galaxy is off by
several hundreds pc from where the bright cluster lie.
The marginal gain in the dispersion in re-computing the
metallicity gradient  from an off-centre position (see previous
section) confirms that this hypothesis seems unlikely.

\subsection{Selection criterion}
That the average metallicity at the centre seems lower than at 1.5~kpc
is hard to explain in
the framework of an inside-out disk formation scenario. We now
discuss the possibility than in the central regions the
metallicity might be higher
than reported in this paper because of a bias in
the \hii\ region selection. As explained in Sect. 3.1
the inclusion of \hii\ regions in our sample requires 
determining of the electron temperature through the detection
of the faint oxygen auroral line. As the metallicity increases
the line becomes so faint as to be detectable only in bright
complexes. The centre of M33 lacks of vigorous star-forming
sites, so the most cooler, metal rich \hii\ regions
have the oxygen auroral line below the detection threshold.
To prove that this might be the case we searched  
the literature for the existing \hii\ region  
spectra inside 1.5 kpc radius, which were not included in our
database because of the undetectable \oiii\ 4363 line.
We found  4 \hii\ regions in the database of 
Magrini et al. (\cite{magrini07a}) 
for which optical spectroscopy is available but no detection
of temperature diagnostic lines. Their names, coordinates,  
galactocentric distances, assumed electron temperatures, and
oxygen abundances from M07a fluxes are shown in Table \ref{tab_cr}. 

\begin{figure}
  \centering
  \includegraphics[angle=-90,width=8cm]{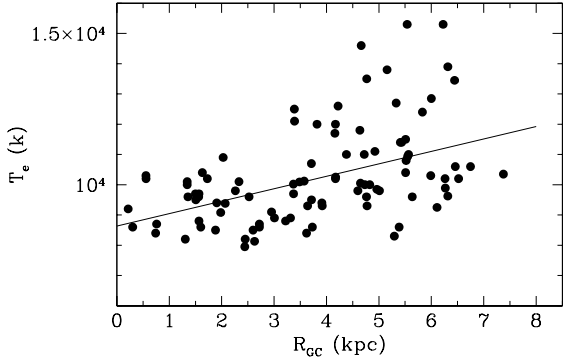}  
    \caption{The radial gradient of \te\ for the \hii\ regions of the  present sample and literature data. The continuous line is a mean least square fit.  }
             \label{fig_temp}%
   \end{figure}

In Fig.\ref{fig_temp} we plot the relationship between the electron temperature  
and the galactocentric distance  for the complete \hii\ region sample.
The weighted  mean least square fit gives a relationship between 
the two quantities
\begin{equation}
T_{e}=(410\pm80) \times R_{GC} +  8600\pm320
\label{eqte}
\end{equation} 
where \te\ is expressed in K and R$_{GC}$ in kpc. 
We thus  derived \te\ for the \hii\ regions of Table \ref{tab_cr}, adopting it 
in their chemical abundance calculations.

\begin{figure}
  \centering
  \includegraphics[angle=-90,width=8cm]{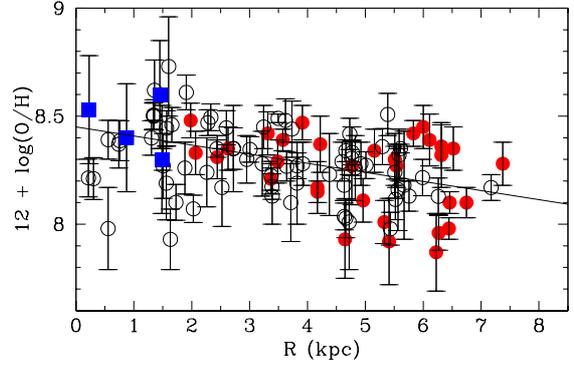}
    \caption{The O/H radial gradient: the symbols for the \hii\ regions with a positive detection 
    of \te\ diagnostic lines are as in Fig. 1, while the filled squares are the regions of Table \ref{tab_cr} for which the 
    \te\ is extrapolated from Eq. \ref{eqte}.  The continuous line is the fit of Eq. 4. }
             \label{Fig_oxy_cr}%
   \end{figure}

Using the intensities by M07a, complemented with upper limit  
measurements of the \oii\ 7320-7330 \AA\,  
when not available in the original paper, we roughly  estimated the  
oxygen abundance of the four
\hii\ regions located within 1.5 kpc. 
Their location in the radial gradient is shown in Fig. 
\ref{Fig_oxy_cr}. The sources clearly lie above the average
metallicity determined in the centre of our database.
It is therefore likely that the adopted 
criterion for source selection, based on the positive detection of
lines for determining the electron temperature, might be responsible 
for the low metallicity derived in the centre.

In the 1.5 kpc central region, there are  
8 \hii\ regions with  
measured electron temperature having oxygen abundances
from $\sim$8 to $\sim$8.5 (see Fig. 1), namely  B0043b  (O/H 8.214),  
B0029  (8.211), B0038b (8.391),
B0016  (7.98), B0079c (8.386), B0027b  (8.37), B0033b   (8.399), B0090  
(8.506).
Their metal abundance and dispersion are consistent with the hypothesis
that we are missing several high-metallicity  
regions within 1.5 kpc of the centre since they only seem to trace
the low metallicity side of the distribution   
present at each given radius.
Similarly, this bias partially explains the different gradient of
the {\it giant} \hii\ regions: we only include metal-rich \hii\
regions (with metallicity above a critical value) in the sample 
if they are very luminous because only these show
detectable temperature diagnostic lines. 

  \begin{table}
\caption{\hii\ regions in the central regions without direct \te measurement.}
\label{tab_cr}
\scriptsize{
\begin{tabular}{llllll}
\hline
\hline
  ID     & RA        & Dec                       &  R$_{GC}$ &  
\te & 12 + log(O/H) \\
           & \multicolumn{2}{c}{J2000.0}       &    (kpc)          & (K)    & dex \\
& (1)     & (2)        & (3)                       & (4)& (5)  \\
BCLMP~93a  & 1:33:52.6 & 30:39:08    & 0.23           &  8700        & 8.53$^{a}$\\
BCLMP~301h  & 1:33:52.6 & 30:45:03     & 1.45         &  9200        & 8.60 \\
BCLMP~4a    & 1:33:59.5 & 30:35:47       &  1.48        &   9200        & 8.33\\
M33-SNR~64 & 1:34:00.1 & 30:42:19       &  0.87         &  8950        & 8.40\\
\hline
\hline
\end{tabular}
a) V\'{\i}lchez et al. \cite{vilchez88} (CC93) adopted \te=6000 k  
obtaining 12 + log (O/H)=9.02
}
\end{table}

\section{The chemical evolution of M33}
\label{sect_model}

The metallicity gradient derived in Sect.\ref{sect_grad} from our sample of
\hii\ regions characterizes the ISM composition in M33 at the present time.
Together with the metallicity gradient from the PN population
(M09), these data allow setting new constraints on current models of
galactic chemical evolution.

The model of M07b, specifically designed for M33, is able to predict
the radial distribution of molecular gas, atomic gas, stars, SFR, and
the time evolution of the metallicity gradient. In the following, we
discuss the modifications needed  to
reproduce the metallicity gradient of \hii\ regions and PNe derived in
this work and in M09.

\subsection{A revised model of chemical evolution}

The multiphase model adopted by M07b follows the formation and
destruction of diffuse gas, clouds, and stars, by means of the simple
parameterizations of physical processes (e.g. Ferrini et al.
\cite{ferrini92}).  In particular, the SFR is the result of two
processes: cloud-cloud interactions (the dominant process) and star
formation induced by the interaction of massive stars on molecular
clouds.  At variance with other models, the relationship between the SFR surface density, $\Sigma_{SFR}$,
and the molecular 
gas surface density, $\Sigma_{H_2}$, (the so-called Schmidt law) 
is thus a by-product of the model and is not assumed.
In general, the relation between the surface density of
total gas and SFR has a slope of $1.4\pm 0.1$, but the slope can vary 
from galaxy to galaxy (Kennicutt \cite{ken98}).
In the particular case of M33, a tight correlation exists between 
the SFR, measured from the FUV emission, and the surface density 
of molecular gas has a well-defined slope (Verley et al. 2009), 
\begin{equation}
\Sigma_{SFR} = A  \Sigma^{1.1\pm0.1}_{H_2}.
\label{eq_sfl}
\end{equation}

According to M07b, the best model for M33 (the so-called {\em
accretion} model, with almost constant infall) suggests a long lasting
phase of disk formation resulting from a continuous accretion of
intergalactic medium during the galaxy lifetime.  We refer to M07b for
the general description of the model and of the adopted parameters for
M33. Here we concisely describe the model and only the
updated processes and equations in detail.

The galaxy is divided into $N$ coaxial cylindrical annuli with inner
and outer galactocentric radii $R_i$ ($i=1,N$) and $R_{i+1}$,
respectively, mean radius $R_{i+1/2}\equiv (R_i+R_{i+1})/2$ and height
$h(R_{i+1/2})$.  Each annulus is divided into two {\em zones}, the {\em
halo} and the {\em disk}, each made of diffuse gas $g$, clouds $c$,
stars $s$, and stellar remnants $r$.  The {\em halo} component includes
the primordial baryonic halo but also the material accreted from the
intergalactic medium.

At time $t=0$, all the baryonic mass of the galaxy is in the form of
diffuse gas in the halo. At later times, the mass fraction in the
various components is modified by several conversion processes: diffuse
gas of the halo falls into the disk, diffuse gas is converted into
clouds, clouds collapse to form stars and are disrupted by massive
stars, and stars evolve into remnants and return a fraction of their mass
to the diffuse gas. In this framework, each annulus  evolves
independently (i.e. without radial mass flows) keeping its total mass
(halo $+$ disk) fixed from $t=0$ to $t_{\rm gal}=13.6$~Gyr.

The disk of mass $M_D(t)$ is formed by continuous
infall from the halo of mass $M_H(t)$ at a rate
\begin{equation}
\frac{{\rm d}M_D}{{\rm d}t}=fM_H,
\label{eqm1}
\end{equation}
where $f$ is a coefficient proportional to the inverse of the infall
timescale. 
Clouds condense out of diffuse gas at a rate $\mu$ and are disrupted by
cloud-cloud collisions at a rate $H^\prime$,
\begin{equation}
\frac{{\rm d}M_c}{{\rm d}t}=\mu M_g^{3/2}-H^\prime M_c^2,
\label{eqm2}
\end{equation}
where $M_g(t)$  is the mass fraction (with respect to the total baryonic 
mass of the galaxy) of diffuse gas, and $M_c(t)$ was defined above.  

Stars form  by cloud-cloud interactions at a rate $H$ and by the
interactions of massive stars with clouds at a rate $a$,
\begin{equation}
\frac{{\rm d}M_s}{{\rm d}t}=H M_c^2+aM_sM_c-DM_s,
\label{eqm3}
\end{equation}
where $M_s(t)$ is the mass fraction in stars, $D$  the stellar death
rate  and $M_c(t)$  as above.  All rate
coefficients of the model are assumed to be independent of time but
are functions of the galactocentric radius $R_{GC}$ (cf. M07b for their values
and radial behaviors).

\subsection{The  choice of the stellar yields and the IMF}

The model results are sensitive to the assumed  stellar yields. For low- and intermediate-mass stars
($M<8$~$M_\odot$), we used the yields by Gavil\'an et
al.~(\cite{gavilan05}). 
The yields of Marigo (\cite{marigo01}) give comparable results without 
any appreciable difference in the computed gradients  of chemical
elements produced by intermediate mass stars, such as N with respect to the gradients 
computed with the yields by Gavil\'an et
al.~(\cite{gavilan05}). 
For stars in the mass range $8~M_\odot < M < 35~M_\odot$ we adopt the yields by Chieffi
\& Limongi~(\cite{chieffi04}). The yields
of more massive stars are affected by the considerable uncertainties
associated with different assumptions about the modelling of processes
like convection, semi-convection, overshooting, and mass loss. Other
difficulties arise from the simulation of the supernova explosion and
the possible fallback after the explosion, that strongly influences the
production of iron-peak elements. It is not surprising then that the
results of different authors (e.g. Arnett~\cite{arnett95}; Woosley \&
Weaver~\cite{woosley95}; Thielemann et al.~\cite{thielemann96}; Aubert
et al.~\cite{aubert96}) disagree in some cases by orders of magnitude. 
In our models, we estimate the yields of stars in the
mass range $35~M_\odot <M<100~M_\odot$ by linear extrapolation of the
yields in the mass range $8~M_\odot < M < 35~M_\odot$.

Another important ingredient in the chemical 
evolution model is the initial mass function (IMF).
Several works support the idea that the IMF is universal in space and
constant in time (Wyse~\cite{wyse97}; Scalo~\cite{scalo98};
Kroupa~\cite{kroupa02}), apart from local fluctuations. 
There are several parameterizations of
the IMF  (see e.g. Romano et al.~\cite{romano05} for 
a complete review), starting from, e.g., 
Salpeter~(\cite{salpeter55}),  
Tinsley~(\cite{tinsley80}), Scalo~(\cite{scalo86}), Kroupa et
al.~(\cite{kroupa93}), Ferrini et al.~(\cite{ferrini90}), Scalo~(\cite{scalo98}),
Chabrier~(\cite{chabrier03}). 
In the following, we test the possibility that the observed
flat gradients can be explained in terms of a non-standard IMF.
In fact, the magnitude and the slope
of chemical abundance gradients are related to the number of stars in
each mass range, and so to the IMF.  
The goal is to reproduce, if possible,  the flat gradient 
supported by recent observations by only varying the IMF. 

\begin{figure}
\centering
\includegraphics[angle=270,width=8cm]{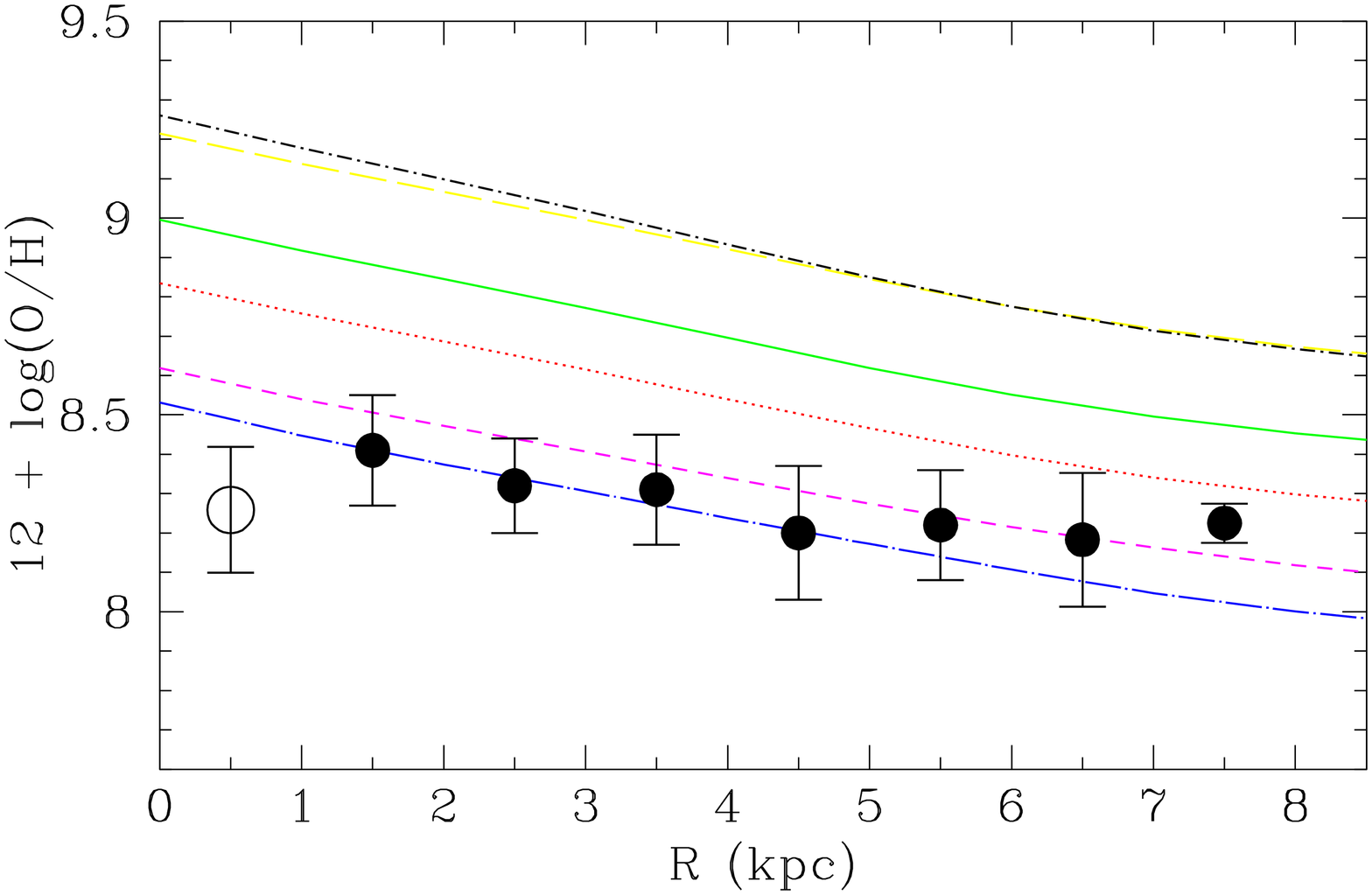}  
\includegraphics[angle=270,width=8cm]{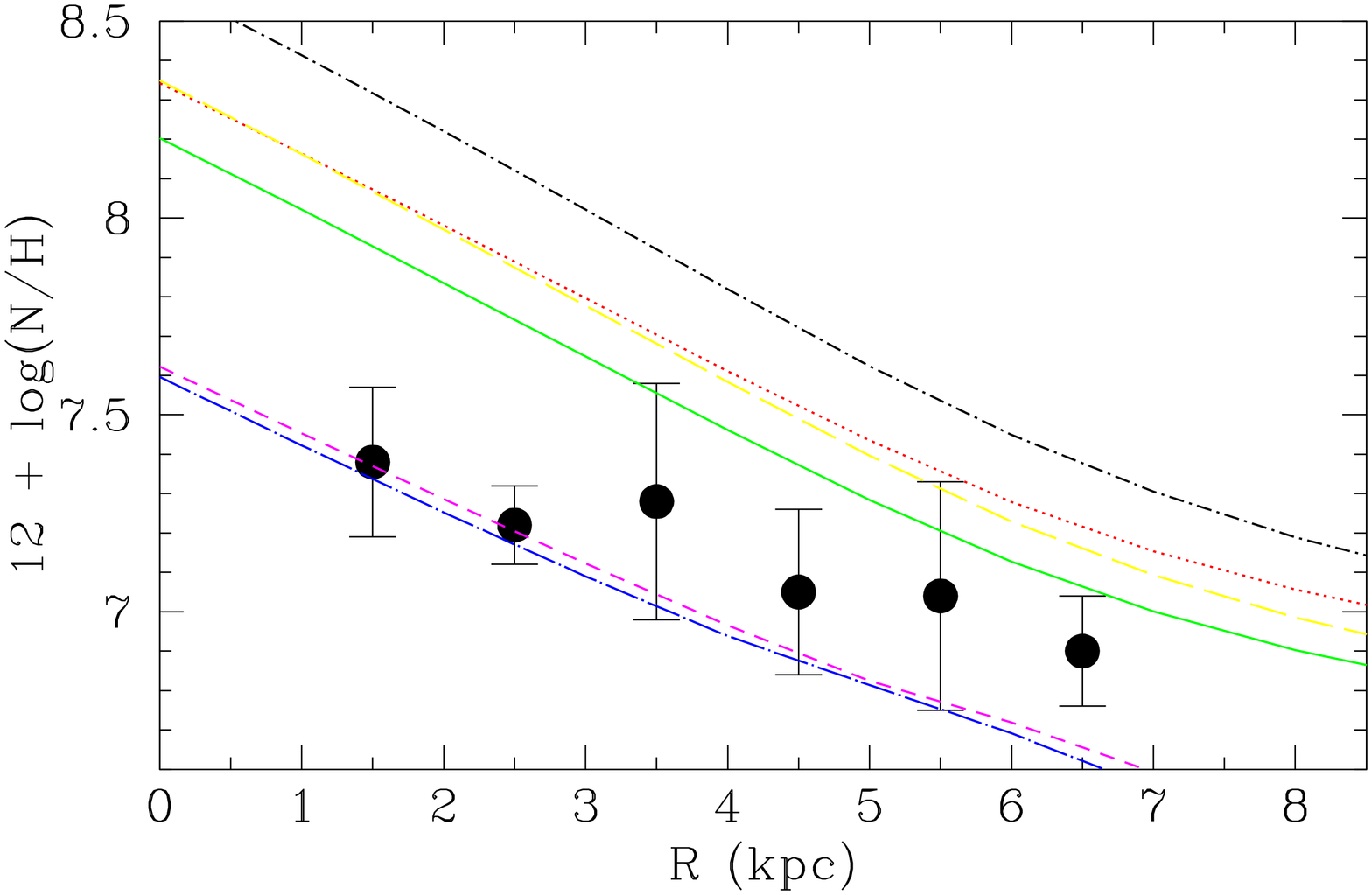}  
\caption{The oxygen (top panel) and nitrogen (bottom panel) radial gradients of M07b with several parameterization of  the IMF: 
Salpeter~(\cite{salpeter55}) (continuous green line), 
Tinsley~(\cite{tinsley80}) (dotted red line),  Scalo~(\cite{scalo86}) (dashed magenta line), Scalo~\cite{scalo98}) (long dashed yellow line), 
Ferrini et al.~(\cite{ferrini90}) (long dash-dotted blue line), Chabrier~(\cite{chabrier03}) (dash-dotted black line).  }
\label{fig_imf}
\end{figure}

In Fig. \ref{fig_imf} we compare the oxygen and nitrogen gradients of
\hii\ regions with present-day abundance profiles from the M07b model.
The choice of the IMF does not affect the slope of  O/H and 
N/H gradients, but simply shifts the abundance profiles
to higher (or lower) values according to the amount of stars that produce oxygen (massive stars) or nitrogen (low- and intermediate-mass stars). 
With the adopted stellar yields, the best fits of the metallicity distribution
of M33 are obtained with the IMF by Ferrini et al.~(\cite{ferrini90})  and  Scalo~(\cite{scalo86}). 
For the revised model we thus adopt the parameterization of Ferrini et al. (1990).
The different slopes of the O/H and N/H gradients are  reproduced fairly well by the chemical 
evolution model as a natural consequence of the different mass ranges of  stellar production, hence timescales,
of these two chemical elements (see Sect.\ref{sec_other} for a comparison with the measured gradients). 

\subsection{Comparison with the observations}
The observational constraints to the model are those described in M07b,
complemented with the O/H and N/H radial gradients of \hii\ regions and PNe
from this work and M09, respectively, and the radial profile of the SFR
determined by Verley et al.~(2009) from far ultraviolet (FUV)
observations corrected for extinction.  For \hii\ regions we used the
gradient derived from the whole population (see Eq. 4), without any
distinction in terms of size and brightness, but excluding the central 1 kpc. 
Giant \hii\ regions might not be representative of the current ISM
abundance due owing their possible  chemical self-enrichment.

\begin{figure}
\centering
\includegraphics[angle=270,width=8cm]{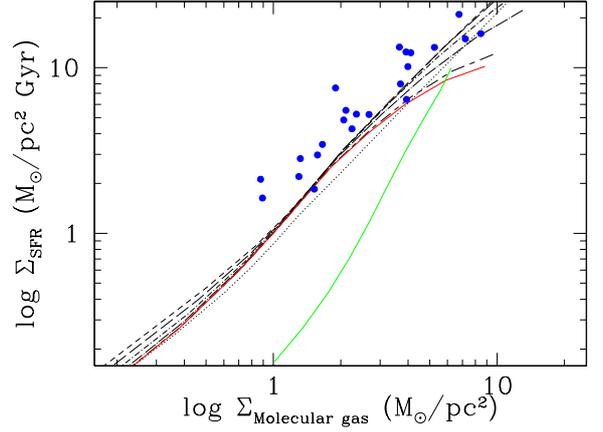}  
\caption{The SFR derived from 
the UV emission corrected for extinction by Verley et al. (2009)
vs. the surface density Schmidt law: filled circles (blue) are the molecular 
gas averaged in bins 1 kpc each (Corbelli \cite{corbelli03}; curves represent the model 
at 0.5 Gyr from the disk formation ({\em dotted curve}), 2 ({\em dashed curve}), 
3 ({\em long-dashed curve}), 5 ({\em dot-dashed curve}), 
8 ({\em long dash-dotted curve}), 12 ({\em long-short dashed curve}), 
and at 13.6~Gyr ({\em solid red curve}); the solid green curve is the model 
by M07b at 13.6 Gyr.  }
\label{fig_sl}
\end{figure}

\begin{figure}
\centering
\includegraphics[angle=270,width=8cm]{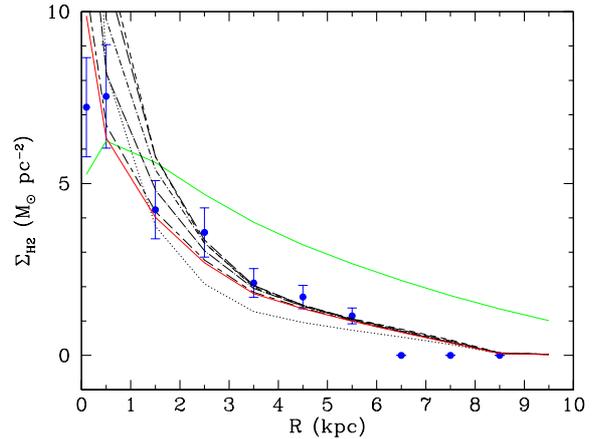}  
\caption{The radial surface density of molecular gas: filled circles (blue) are the observed molecular gas averaged in bins 1 kpc wide. Model curves have the same symbols as in Fig. \ref{fig_sl}. }
\label{fig_mol}
\end{figure}

In Figs. \ref{fig_sl} and \ref{fig_mol} we show the radial surface density of molecular gas (from
the single dish observations of Corbelli~\cite{corbelli03}, averaged
over 1 kpc bins) and the SFR as a function of the surface density of
molecular gas (Verley et al.~2009), respectively.  
Both
figures show the predictions of the model of M07b at
$t=t_{\rm gal}$.  Clearly, the model of M07b in its original formulation
is unable to reproduce these constraints.  We have therefore
considered other parameterizations of the star formation process by
varying the radial dependence of the SFE, represented by the coefficient $H$.

Our experiments show that, to reproduce the
observed behaviour of the radial gas distribution and the observed Schmidt law, 
is necessary to increase the efficiency of star formation at large
radii. This can be accomplished in many ways.
In the previous version of the model (M07b) $H$ decreased with $R_{GC}$ 
to consider a geometrical correction resulting
from how  large galactocentric distances correspond
to larger volumes (see Ferrini et al. \cite{ferrini94}). In the present work
we assume that $H$ is constant with radius, thus implying that
the star formation efficiency increases linearly with
galactocentric radius. In  Sect. \ref{sec_ocon} we describe 
the observational evidence in support this assumption in M33. 

Assuming that $H$  is  spatially constant, we obtain the results 
shown in
Figs.~\ref{fig_sl} and \ref{fig_mol} at $t=0.5$, 2, 3, 5, 8, 12, 13.6
Gyr. The relationship between molecular gas and SFR predicted
by the modified M07b model is more or less constant with time and in
good agreement with the data. The predicted Schmidt law (at the present
time) has an average exponent $\sim 1.2$, in agreement with the
observations, whereas the original M07b model produces a Schmidt law
with a higher exponent, $\sim 2.2$.  The better agreement with the
observations obtained with the revised M07b model suggests that M33 is
more efficient in forming stars than normal local Universe spiral
galaxies, in particular in its outer regions.

\subsection{Flat gradients and SF efficiency}
\label{sec_ocon}

Is there evidence of flat gradients in other galaxies? 
How can they be explained? 
A flat gradient in the outer regions of our Galaxy has been observed by several authors 
(e.g., Yong et al. \cite{yong05}, Carraro et al.~\cite{carraro07}; Sestito et al.~\cite{sestito07}) 
 using different metallicity tracers (e.g., Cepheids, open clusters). 
 However, other tracers, such as PNe, do not show this behaviour, indicating  flat gradients  
 across the whole disk (cf. Stanghellini et al. \cite{stanghellini06}, Perinotto \& Morbidelli \cite{peri06}).
The  outer Galactic plateau might be a phenomenon similar to the flat gradient of M33.  
Since M33  is less massive than the MW, its halo collapse 
phase, responsible for the steep gradient in the inner regions (R$_{GC} <$11-12 kpc)
of our Galaxy (cf. Magrini et al. \cite{m09}), is less marked. 
Thus in M33 the difference between the {\em inner} and {\em outer} gradients is 
less evident than in the MW.  
The flat metallicity distribution of the MW at large radii
has been explained with several chemical evolution models, among them 
those by Andrievsky et al.~(\cite{andri04}), Chiappini et
al.~(\cite{chiappini01}) model C, and  Magrini et al. (\cite{m09}).
Andrievsky et al.~(\cite{andri04}) explain the flat metallicity distribution beyond 11-12 kpc 
when assuming that the SFR is a
combination of two components: one proportional to the gas surface
density, and the other depending on the relative velocity of the
interstellar gas and spiral arms. With these assumptions they explain 
why the breaking point in the slope of the gradient and the 
consequent outer flattening, occur around the co-rotational radius.    
Chiappini et al.~(\cite{chiappini01}) assume two main accretion
episodes in the lifetime of the Galaxy, the first one forming the
halo and bulge and the second one forming the thin disk.  
Their model C is the one able to reproduce a flat gradient in the 
external regions, assuming that there is no threshold in the gas density
during the halo/thick-disk phase, and thus allowing the formation of the
outer plateau from infalling gas enriched in the halo.
Finally, Magrini et al. (\cite{m09}) reproduce the Galactic gradient  thanks to the 
radial dependence of the infall rate (exponentially decreasing with
radius)  and with the radial
dependence of the star and cloud formation processes.
To reproduce a completely flat gradient in the outer regions would, 
however, require additional accretion of gas uniformly falling onto the disk, which would  result in 
inconsistent behaviour of the current SFR. 

{\em Why do we use a radially increasing SFE to reproduce the flat metallicity  gradient of M33?}
M33 in general is quite different from large spiral galaxies in terms of SF. A comparison of the SFR 
to the H$_{2}$ mass shows that M33, like the intermediate redshift galaxies, has a significantly
higher SFE than large local Universe spirals.
There is also observational evidence that  the SFE varies with radius in M33. 
Kennicutt (\cite{ken98b}), Wong \& Blitz (\cite{wong02}), and Murgia et al. (\cite{murgia02}) find a molecular gas depletion
timescale (inversely proportional to the SFE)  that varies radially, decreasing  by a factor $\sim$2-3.
Also Gardan et al. (\cite{gardan07}) from the CO to SFR ratio also find a dependence
of the depletion timescale on radius,  decreasing  of  a factor $\sim$2 over 4 kpc.
These observations prompted us to model the chemical evolution of M33 taking  SF processes 
depending on radius into account. 
The general assumption of our model is that the SF is driven by cloud collisions. Increasing the efficiency 
of this process with radius does not imply that cloud-cloud collisions are more efficient in the outer regions, 
but that additional processes may contribute to the SF in the peripheral regions.

\subsection{Implications for the metallicity gradient and  its evolution}

The introduction of more realistic SF process in M33 has significant
consequences on its metallicity gradient and evolution.  As described
by M09, oxygen modification (destruction or creation) does not occur in
the PN population of M33.  Thus oxygen, the best-determined element in
nebular optical spectroscopy, can be safely used as a tracer of the ISM
composition at the epoch of the formation of PN progenitors. Most PNe
in M33 are older than 0.3 Gyr, and probably much older, up to an age of
10 Gyr.

The metallicity gradient of PNe, including only those with progenitor stars older than 
0.3 Gyr and excluding PNe located in the first  kpc
from the centre, is 
\begin{equation}
12 + {\rm log(O/H)} = -0.040 (\pm 0.014) ~  {\rm R_{GC}} + 8.43 (\pm0.06). 
\end{equation}
As discussed above, the whole sample of \hii\ regions is representative
of the current ISM composition. 
The \hii\ region and PN gradients in the same radial region, 1 kpc $\lesssim R_{GC} \lesssim$ 8 kpc, 
are indistinguishable within the
uncertainties in their slopes, with a slightly  translation, $\sim$0.1 dex, towards higher metallicity in the \hii\ region sample.   
This means that very little evolution of metallicity has occurred in the past few Gyr.

\begin{figure}
\centering
\includegraphics[angle=0,width=10cm]{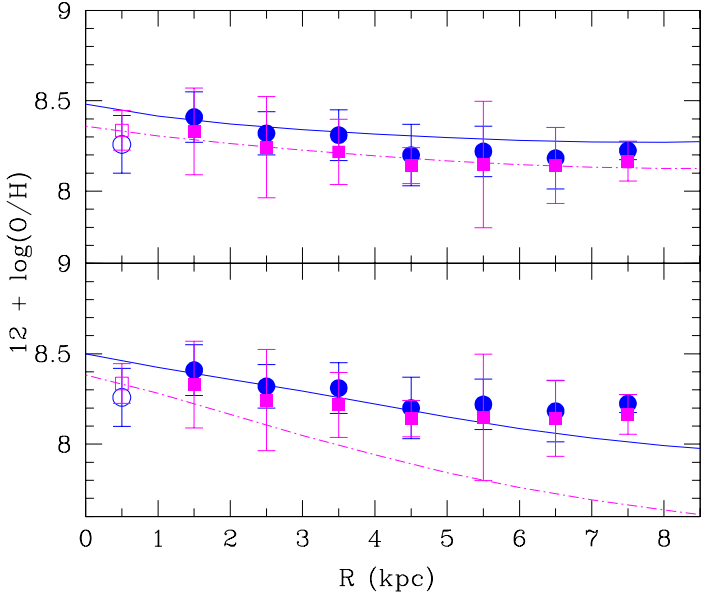}  
\caption{The O/H radial gradient evolution.
Top panel: filled circles (blue) are the \hii\ region 
oxygen abundances from the dataset described in Sect.\ref{sect_mmt}, 
averaged over radial bins, each 1 kpc wide; filled squares (magenta) are 
averaged non-Type I PN oxygen abundances by M09. 
The gradient 
5 Gyr ago ({\em  dashed curve}) and now (continuous curve) as predicted by the 
present model.
Bottom panel: the same as in the top panel but for the {\em accretion} model by M07b.}
\label{fig_grad}%
\end{figure}

In Figure \ref{fig_grad} we show the observed metallicity gradient of M33 
together with  the results of the present model (top panel) and  of the original 
model of M07b (bottom panel). For a better comparison,
the oxygen abundances
of the samples of \hii\ regions and disk PNe have been averaged
over bins of 1 kpc each. This allows us to highlight the translation 
towards the lower metallicity of the PNe. The curves 
correspond to the present  and to 5 Gyr ago (8.6 Gyr from
the disk formation, assuming an age of 13.6 Gyr for M33), i.e.,  approximately the
average period of formation of the PN progenitors. 
the present
model predicts a higher SF in the outer regions, reproducing the
observed flatness of the oxygen gradient at large radii better than the M07b
model.  In addition, the evolution of the metallicity
gradient predicted by the new model is consistent with the
observations: a negligible change in the slope and a small
translation to higher metallicities. Finally, we see that the assumed  
dependence of the SFE with galactocentric radius,
required to reproduce the metallicity evolution of M33, 
takes place always as in M07b with a slow building up of 
the disk of M33  by accretion from an intergalactic
medium or halo gas. 

For completeness we also show the N/H radial gradient of the sample of \hii\ regions. 
For this element the temporal evolution cannot be inferred using PN abundances, 
since PNe modify their nitrogen composition during their lifetime. 
The nitrogen abundances have been averaged
over bins of 1 kpc each.
The observed N/H gradient is significantly steeper than the oxygen gradient, 
as predicted by the model. 

\begin{figure}
\centering
\includegraphics[angle=0,width=10cm]{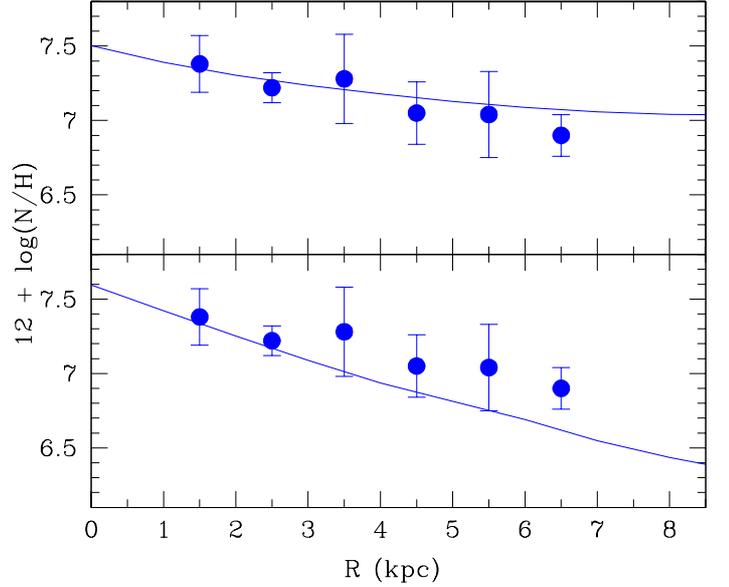}  
\caption{The N/H radial gradient.
Top panel: filled circles (blue) are the \hii\ region 
nitrogen abundances from the dataset described in Sect.\ref{sect_mmt}, 
averaged over radial bins, each 1 kpc wide.
The gradient at the  present time  as predicted by the 
present model is the continuous curve.
Bottom panel: the same as in the top panel but for the {\em accretion} model by M07b.}
\label{fig_grad2}%
\end{figure}

The time evolution of the metallicity gradient is strictly correlated to the 
inside-out growth of the disk. 
In a recent work, Williams et al. (\cite{williams09}) presented the resolved photometry of four fields located at different 
galactocentric distances, from 0.9 to 6.1 kpc. 
Their photometry provides a detailed census of  stellar populations and their ages  at different radii.
They find that the percentage of the stellar mass formed prior to {\em z}= 1 ($\sim$8 Gyr ago) changes
from $\sim$71\% in the innermost field at 0.9 kpc from the centre
to $\sim$16\% in the outermost SF region at 6.1 kpc. 
We thus compare these results with the cumulative SF predicted by our model 
at different ages. The agreement is good in the central field, where model and observations  both have  
about 70\% of the stellar mass already formed 8 Gyr ago. Also at 2.5 kpc, 
corresponding to their second field, the observed 
stellar mass formed  prior to {\em z}= 1 is about 50\% and the model results give 
47\%. 
In the outermost fields, at 4.3 and 6.1 kpc, the agreement is not as good, 
since the observations show that 20\% and 16\% of stars formed before 8 Gyr ago, while 
the model predicts  $\sim$44\% for both fields.  
With the  model by M07b, the situation is almost unvaried in the inner regions, 
while in the outer regions there is a slightly lower percentage of stars formed before 8 Gyr ago, $\sim$33\%. 
The recent evolution of the outermost SF regions of M33
can be explained by an accretion of material in the peripheral regions
at recent epochs.    
In our model, the scalelength of the disk is constant, and the accretion in the outer 
regions is a continuous process. Thus, we are 
able to reproduce the integrated SF at any radius and its by-product, the metallicity, 
but not to reproduce  its temporal behaviour if it is dominated by stochastic events, such as, e.g.,  
sporadic massive accretion of gas and tidal 
interactions.  
Finally, outside the SF area of the disk, the age of populations of the outer disk/halo show an age increase 
with radius (Barker et al. \cite{barker07a}, \cite{barker07b}). This agrees well with our model results: there is 
a slightly larger number of old stars beyond 8 kpc than in the peripheral part of the SF disk.

\section{Summary and conclusions}
\label{sect_conclu}
We have studied the chemical evolution of M33  by means of new spectroscopic observations 
of \hii\ regions, together with literature data of \hii\ regions and PNe.
We derived the radial oxygen gradient in the range from 1 to $\sim$8 kpc. Its  slope is -0.044$\pm$ 0.009
dex kpc$^{-1}$.  
We excluded the central 1 kpc region from the gradient because 
of its low metallicity. 
In fact, the 2D metallicity map is off-centre, with a peak in the southern
arm at 1-2 kpc from the centre. We measured the highest metallicity gradient in 
bright regions which are not located in the centre of M33.  
We explained this effect  with a bias in the measurement of nebular 
abundances towards low metallicity in the central regions and/or with SF bursts which 
had not had time to  mix azimuthally.
At a galactocentric distance of about 1.5 kpc, the spread of metal abundances is much greater than 
the dispersion along the average gradient, and it is possibly related to the presence of a bar. 

We analysed and discussed the metallicity gradient of {\em giant} \hii\ regions, i.e. the largest regions with high surface brightness. 
We find it steeper than the average of the sample  gradient, in agreement with the early spectroscopic 
studies of \hii\ regions in M33, 
based mainly on the brightest objects. 

We compared the metallicity gradients of \hii\ regions and of PNe, obtained with the same set of observations and 
analysis techniques. We find a substantially unchanged slope of the gradient, and an overall increase in metallicity with time.
We can explain the slow evolution of the metallicity gradient from the present to the 
birth of the PN progenitors  with a chemical evolution {\em accretion} model, as in M07b, if the
SFE  is higher at larger radii.

\appendix
\section{Comparison with the literature}
\label{sec_comp}

To assess the quality of the MMT observations we
compared the oxygen abundance in  the control sample with those available in the literature.
Oxygen is in fact the most widely used element in emission-line optical spectroscopy 
because it gives accurate results for determining the metallicity. This is because
{\em i}) many ionization stages are present in the optical range (\oi, \oii, \oiii), and the 
corrections for unseen ionization stages are often not needed ; {\em ii}) a direct measurement 
of its electron temperatures,  \teoiii\ and \teoii, is possible.
For each source in the control sample, we show in Table \ref{tab_hii_comp} the oxygen abundance 
determined by different authors    
(Rosolowsky \& Simon \cite{rs08},  Magrini et al. \cite{magrini07a}, Crockett et al. \cite{crockett06}, 
Vilchez et al. \cite{vilchez88}, Kwitter \& Aller \cite{kwitter81}) and in this paper. 
In columns  2 to 6 we report the literature O/H and our measurements in column 7. 
For \hii\ regions with multiple determinations we computed the weighted average of all measurements. 
The agreement is reasonably good for all sources, since our values are consistent, within the errors, 
with those available in the literature. 
The average O/H abundance of the 14 \hii\ regions agrees. The average dispersion 
between our values and previous determinations is only 0.09 dex.

\begin{table*} 
\label{tab_hii_comp}
\caption{Comparison of oxygen measurements of our control sample with  literature O/H.}
\scriptsize{
\begin{tabular}{lllllll}
\hline\hline
ID   					&\multicolumn{6}{c}{12 +log(O/H)}                           \\
              					&M07a             		& RS08      & C06 & KA81  & V88      &pw  \\
(1)						&(2)								&(3)			&(4)	&(5)		&(6)		 &(7)\\					
\hline
LGCHII2             		    &   8.25$\pm$0.06   &    			&		   &		  &				&    8.10$\pm$0.05 \\
LGCHII3             		    &   8.24$\pm$0.05   &				&		   &		  &    			&    8.42$\pm$0.06 \\
BCLMP289            		&   8.25$\pm$0.13  & 				&  		   &           & 			&    8.35$\pm$0.12 \\
BCLMP218            		&   8.25$\pm$0.05  &	8.16$\pm$0.06   &&           &             &    8.17$\pm$0.12 \\
MA1                 			&   						&				&8.24$\pm$0.06&&          &    8.28$\pm$0.15  \\
BCLMP290            		&   						&				&8.21$\pm$0.06 &&    		&    8.37$\pm$0.13 \\
IC132               			&   8.08$\pm$0.04   &    			&		   &  		    &			&  7.98$\pm$0.05    \\
BCLMP45             		&   8.49$\pm$0.04   &    		     & 		   &			&			&   8.48$\pm$0.08 \\
BCLMP670            		&   8.28$\pm$0.07   &				&   	   &			&	         &    8.29$\pm$0.07\\
MA2                 			&   						&8.33$\pm$0.08&  &     8.38	&8.44$\pm$0.15    &    8.31$\pm$0.10    \\
BCLMP691            		&   						&				&8.26$\pm$0.02&&      	&    8.42$\pm$0.06 \\
IC131		            		&   						&				&				&&  8.43$\pm$0.03    	&    8.47$\pm$0.08 \\
IC133       			     		&   						&	8.23$\pm$0.05			&&&      	&    8.27$\pm$0.08 \\
BCLMP745            		&   						&				&8.07$\pm$0.10&&      	&    7.93$\pm$0.10 \\
%
\hline
\hline
\end{tabular}}\\
M07a--Magrini et al. \cite{magrini07a}, RS08--Rosolowsky \& Simon \cite{rs08}, 
C06--Crockett et al. \cite{crockett06}, 
KW81--Kwitter \& Aller \cite{kwitter81}, V88--Vilchez et al. \cite{vilchez88});  present work (pw).
\end{table*}

\begin{acknowledgements}
We warmly thank J. Vilchez for interesting discussion on the argument. 
We thank R. Walterbos for allowing us to  use  his \ha\ calibrated map of M33. 
We thank D. Fabricant  for making 
Hectospec  available to the community, and
 the Hectospec instrument team and MMT staff for their expert help in
preparing and carrying out the Hectospec observing runs. 
We thank N. Caldwell, D. Ming, and their team for the help during the data reduction. 
L.S. acknowledges the hospitality of the Observatoire de Paris where a part of this work 
was developed. 

\end{acknowledgements}

\end{document}